\documentclass[aps,prb,twocolumn,showpacs,superscriptaddress,groupedaddress]{revtex4}  
\usepackage{amsmath}
\usepackage{amssymb}
\usepackage[pdftex]{graphicx}
\usepackage{amsmath}
\usepackage{epsfig}
\usepackage{helvet}
\usepackage{amssymb}
\usepackage{color}
\usepackage [english]{babel}
\usepackage [autostyle, english = american]{csquotes}
\MakeOuterQuote{"}
\DeclareGraphicsExtensions{.pdf,.png,.jpg}

\newcommand{\bea}{\begin{eqnarray}}
\newcommand{\eea}{\end{eqnarray}}

\newcommand{\nn}{\nonumber}

\newcommand{\bra}[1]{\mbox{$\langle #1 |$}}
\newcommand{\ket}[1]{\mbox{$| #1 \rangle$}}
\newcommand{\braket}[2]{\mbox{$\langle #1 | #2 \rangle$}}

\newcommand{\GS}{\mbox{\tiny GS}}

\begin{document}

\title{The iPEPS algorithm, improved: fast full update and gauge fixing}

\author{Ho N. Phien}  \affiliation{Centre for Health Technologies, Faculty of Engineering and Information Technology, University of Technology Sydney, Sydney 2007, Australia}
\author{Johann A. Bengua}  \affiliation{Centre for Health Technologies, Faculty of Engineering and Information Technology, University of Technology Sydney, Sydney 2007, Australia}
\author{Hoang D. Tuan}  \affiliation{Centre for Health Technologies, Faculty of Engineering and Information Technology, University of Technology Sydney, Sydney 2007, Australia}
\author{Philippe Corboz} \affiliation{Institute for Theoretical Physics, University of Amsterdam, Science Park 904 Postbus 94485, 1090 GL Amsterdam, The Netherlands}
\author{Rom\'an Or\'us} \affiliation{Institute of Physics, Johannes Gutenberg University, 55099 Mainz, Germany}
\begin{abstract}

The infinite Projected Entangled Pair States (iPEPS) algorithm [J. Jordan \emph{et al},  PRL {\bf 101}, 250602 (2008)] has become a useful tool in the calculation of ground state properties of 2d quantum lattice systems in the thermodynamic limit. Despite its many successful implementations, the method has some limitations in its present formulation which hinder its application to some highly-entangled systems. The purpose of this paper is to unravel some of these issues, in turn enhancing the stability and efficiency of iPEPS methods. For this, we first introduce the  \emph{fast full update} scheme, where effective environment and iPEPS tensors are both \emph{simultaneously} updated (or evolved) throughout time. As we shall show, this implies two crucial advantages: (i) dramatic computational savings, and (ii) improved overall stability.  Besides, we extend the application of the \emph{local gauge fixing}, successfully implemented for finite-size PEPS [M. Lubasch, J. Ignacio Cirac,  M.-C. Ba\~{n}uls, PRB {\bf 90}, 064425 (2014)], to the iPEPS algorithm. We see that the gauge fixing not only further improves the stability of the method, but also accelerates the convergence of the alternating least squares sweeping in the (either "full" or "fast full") tensor update scheme. The improvement in terms of computational cost and stability of the resulting "improved" iPEPS algorithm is benchmarked by studying the ground state properties of the quantum Heisenberg and transverse-field Ising models on an infinite square lattice. 

\end{abstract}

\maketitle

\section{Introduction}
In the last years, tensor networks have emerged as a powerful tool to understand quantum many-body systems.~\cite{RevTN} From the point of view of numerical simulations there have been a number of novel algorithms developed, whose inner workings are deeply rooted in the theory of quantum entanglement. In all these algorithms quantum many-body states are conveniently represented by tensor networks that efficiently capture the natural structure of quantum correlations in the system (such as, e.g., the  so-called \emph{entanglement area law}.\cite{arealaw1,arealaw2,arealaw3,arealaw4,arealaw5}). Many important classes of states can be accurately approximated by a tensor network with a number of parameters that depends only polynomially on the size of the system. 
Such properties, among others,  have enabled tensor network methods to break the \emph{curse of dimensionality}, namely, the fact that the Hilbert space dimension of a quantum many-body system is exponentially large in the number of particles. Thanks to this,  it is now possible to efficiently simulate  quantum many-body systems by targeting the relevant tiny corner of quantum states (e.g., those satisfying an area-law) inside of the exponentially-large Hilbert space. \cite{Vidal1,Vidal2, Frank}

The so-called matrix product state (MPS) \cite{Fannes1,Ostlund1} is a typical tensor network ansatz representing the state of 1d gapped quantum lattice systems. It is also well known that MPS is the class of variational wave functions at the root of the density matrix renormalization group (DMRG) method,\cite{White1, White2} widely used in the study of 1d systems. \cite{Takasaki1, schollwock1, Perez1, Ian1, Ian2, Greg, schollwock2} Subsequently, MPS methods to study time-evolution of 1d systems have also been put forward, such as time evolving block decimation (TEBD),~\cite{Vidal1,Vidal2} time-dependent DMRG,~\cite{tdDMRGWhite,tdDMRGSchollwock} and more recently algorithms based on the time-dependent variational principle.~\cite{TDVP} One way of generalizing MPS methods to higher dimensional systems \footnote{Other generalizations are possible, using, e.g., tree tensor networks (TTN) \cite{ttn}Êand the multi-scale entanglement renormalization ansatz (MERA) \cite{mera}.} is using projected entangled pair states (PEPS), \cite{Verstraete1, Murg1, Verstraete2}, sometimes also called tensor product states (TPS).~\cite{Nishino2, Nishino3, Gendiar1} Ê There has been a lot of progress both in conceptual and algorithmic developments for PEPS, see, e.g., Refs~\onlinecite{Levin1,Jordan1,Roman2,Jiang1,Xie1,Philippe1,Wang1,Pizorn1,Li1,Xie2,Romancano1,Romancano2,Lubasch1,Wang11, Lubasch2,Phiencano1, Bauer}. From the numerical perspective, PEPS have been used to study ground state properties as well as dynamics of 2d lattice systems, both of finite and infinite size. Moreover, motivated by the success of 1d methods in the thermodynamic limit such as iTEBD,~\cite{Vidal3, Roman1} the so-called infinite-PEPS (iPEPS) algorithm~\cite{Jordan1,Roman2} was put forward to study infinite-size 2d quantum lattice systems. 

So far, the iPEPS algorithm has been quite successful in studying ground state properties of a growing number of 2d quantum lattice systems (see, e.g., Ref.~\onlinecite{RevTN} and references therein). In general terms, results obtained by using iPEPS can be competitive when compared to the ones derived from quantum Monte Carlo.~\cite{Roman2,Bauer1} And what is more important, the iPEPS algorithm is not hampered by the \emph{sign} problem (unlike quantum Monte Carlo) when studying fermionic and frustrated spin systems. Recent applications of iPEPS to such systems include calculations for the $t-J$ model of fermions on the square,~\cite{Philippe1, Philippe3,Philippe5} and honeycomb lattices,~\cite{Gu} as well as the $J_1-J_2$ frustrated Heisenberg model on the square lattice,~\cite{Wang} the Shastry-Sutherland model,~\cite{Philippe5b} and the Kagome Heisenberg antiferromagnet.~\cite{KHA} 

One of the main drawbacks of the iPEPS algorithm is its high computational cost as a function of the bond dimension $D$ which controls the accuracy of the method. This is particularly true when trying to obtain good accuracies in physical regimes where entanglement is large (e.g., close to a quantum critical point, or in the presence of many nearly-degenerate quantum states), which requires a large $D$. The computational bottleneck of the method is the calculation of the so-called \emph{effective environment}, i.e., the effective description of the tensor network surrounding a given site. Technically, effective environments can be computed using different approaches, such as   TRG/SRG and HOTRG/HOSRG,~\cite{Levin1,Xie1,Zhao1,Xie2} iTEBD,~\cite{Jordan1} corner transfer matrices (CTM),~\cite{Baxter11,Nishino1,Nishino1_1,Roman2,Philippe4,RomanNewCTM1} or more recently the tensor network renormalization method.~\citep{GlenRenormT} Independently of the chosen approach for their calculation, accurate effective environments are required for the so-called \emph{full update}~(FU), which is the accurate scheme proposed to find the iPEPS tensors throughout a (imaginary) time evolution. Because of the large computational cost only relatively small bond dimensions can be afforded when using the FU scheme.
%

Alternative update schemes have been developed trying to overcome this problem, but only with partial success. A very popular approach is the so-called \emph{simple update} (SU),~\cite{Jiang1, Li1} which relies on a mean-field approximation of the effective environment, thus being  very efficient. Such a scheme allows to reach large bond dimensions in the iPEPS but, not surprisingly, it does not produce accurate results when systems are very strongly-correlated. Intermediate approaches interpolating between the FU and the SU schemes have also been put forward as an alternative.~\cite{Lubasch1, Wang11}

So, here is the dilemma: accurate update schemes like the FU are too costly, whereas efficient  update schemes such as the SU are not accurate enough. The question then is: can one somehow "accelerate" the FU, making it more efficient while keeping its accuracy? 

In this paper we give a positive answer to this question. We do this by constructing an update scheme, which we call "fast full update" (FFU), that significantly reduces the computational cost of iPEPS algorithms while still being accurate. More specifically, in this new strategy at every  time step the tensors of the effective environment are updated by a single iteration step (in a sense to be made specific later) and simultaneously with those of the iPEPS. Importantly, we find that applying this strategy to iPEPS algorithms not only reduces the computational cost by a large factor (as expected), but also contributes to stabilize the algorithm. 
The reason for this is that the successively updated environment helps to maintain the compatibility between the related tensors throughout the time evolution, as we shall explain later.

In addition, we show that incorporating the \emph{local gauge fixing} scheme proposed in Ref.~\onlinecite{Lubasch2} (and successfully applied to finite PEPS) can make the iPEPS algorithm even faster and more stable. As described in Ref.~\onlinecite{Lubasch2}, the idea for the gauge fixing of PEPS tensors is inspired by the case of MPS, for which tensors can always be represented in a \emph{canonical form} during their update by means of local gauge fixing. In the canonical form many of the tensor manipulations of an MPS get simplified (or directly canceled out) implying a much better conditioning and stability of related algorithms. However, unlike in the MPS case, there is no exact canonical form for PEPS in the same sense. A recent attempt along this direction is the so-called  \emph{quasi-canonical form} for iPEPS.~\cite{Phiencano1,Romancano1,Romancano2} This has been shown to lead to some computational advantages, but unfortunately does not fully capture the effect of quantum correlations spreading throughout the 2d lattice. A different approach was considered in Refs.~\onlinecite{Phiencano1,Lubasch2}, where it was shown that by considering the effect of the entire 2d lattice, a local gauge choice of the tensors can also produce a well-conditioned environment, which in turn improves the stability of the subsequent calculations. Here we apply the same local gauge fixing as in Ref.~\onlinecite{Lubasch2} to the iPEPS algorithm and show that it not only improves the stability, but also accelerates the convergence of the alternating least squares sweeping in the tensor update scheme. 

To show the validity of these approaches, we provide benchmarking calculations for the "improved" iPEPS algorithm with the two "improvements" mentioned above (FFU + gauge fixing). In particular, we analyse the computational cost and the stability of the algorithm, for ground-state calculations of the Heisenberg and transverse-field Ising models on an infinite square lattice. We shall see quantitatively that the "improvements" both accelerate and stabilize the overall numerical calculations. 

The paper is structured as follows. Some background material on PEPS, iPEPS, and the iPEPS algorithm is presented in Sec. II. In Sec.~III we introduce the improvements (FFU and gauge-fixing). Benchmarking calculations are presented in Sec. IV. Finally, conclusions are presented in Sec.~V. 

\section{\label{secII} Background}

\subsection{PEPS and iPEPS}
\subsubsection{Generalities}
For completeness, we briefly review the notation and fundamental properties of projected entangled pair states (PEPS).~\cite{Verstraete1, Murg1, Verstraete2} To this end, let us consider a 2d quantum lattice system consisting of $N$ sites, each of which is described by a local Hilbert space $\mathbb{C}^{d}$. The full Hilbert space of the system is thus $\mathcal{H} = (\mathbb{C}^{d})^{\otimes N}$. As the dimension of the full Hilbert space grows exponentially with the size of the system, the problem quickly becomes intractable already for moderately-low values of $N$. In order to avoid this curse of dimensionality, an option is to use a PEPS to represent a pure state of the system. Generally speaking, a PEPS is a state defined by a 2d lattice of interconnected tensors, i.e.,
\bea
\ket{\Psi} &=& \sum_{\{s_{\vec{r}_{i}}\}_{i=1}^{N}}^{d} F(A^{[\vec{r}_{1}]}_{s_{\vec{r}_{1}}},\ldots, A^{[\vec{r}_{N}]}_{s_{\vec{r}_{N}}})\ket{s_{\vec{r}_{1}},\ldots,s_{\vec{r}_{N}}},\nn\\
\label{eqC4_1}
\eea
where $\ket{s_{\vec{r}_{i}}}$ is the local basis of the site located at $\vec{r}_{i} = (x_{i},y_{i})$. Depending on the geometry of the lattice pattern, the tensor $A^{[\vec{r}_{i}]}_{s_{\vec{r}_{i}}}$ at each lattice site $\vec{r}_{i}$ contains $n_{\vec{r}_{i}}$ \emph{bond} indexes taking up to ${D}$ values ($n_{\vec{r}_{i}}$ is typically the number of nearest neighbours of the lattice site $\vec{r}_{i}$) and a \emph{physical} index taking up to $d$ values. The operation $F$ contracts all the tensors $A^{[\vec{r}_{i}]}_{s_{\vec{r}_{i}}}$ along the bond indices. Conventionally, $D$ is usually referred to as the \emph{bond dimension}, which plays the role of a parameter quantifying both the size of the tensors in the PEPS, and also the amount of entanglement in the wave function.~\footnote{${D}=1$ is a product state, whereas larger $D>1$ correspond to entangled states. The larger $D$ is, the larger the amount of entanglement in the wave function is, as quantified by the prefactor to the entanglement area-law.~\cite{RevTN}} Thus, the larger $D$, the better the PEPS can represent the state of the system (since there are more variational parameters). As an example, in Fig.~\ref{PEPS}(a) we illustrate a graphical representation of the PEPS for a $5\times 5$ square lattice. In this case the amount of complex parameters describing the PEPS quantum state is $\mathcal{O}(NdD^{4})$, thus polynomial in $N, D$ and $d$, in contrast with the exponential dependence on $N$ for an arbitrary state in the Hilbert space. For the sake of simplicity, from now on we shall always assume that we have a 2d square lattice.~\footnote{Other 2d lattices can also be easily considered.}

\subsubsection{Properties}
\begin{figure}
	\includegraphics[width =\columnwidth]{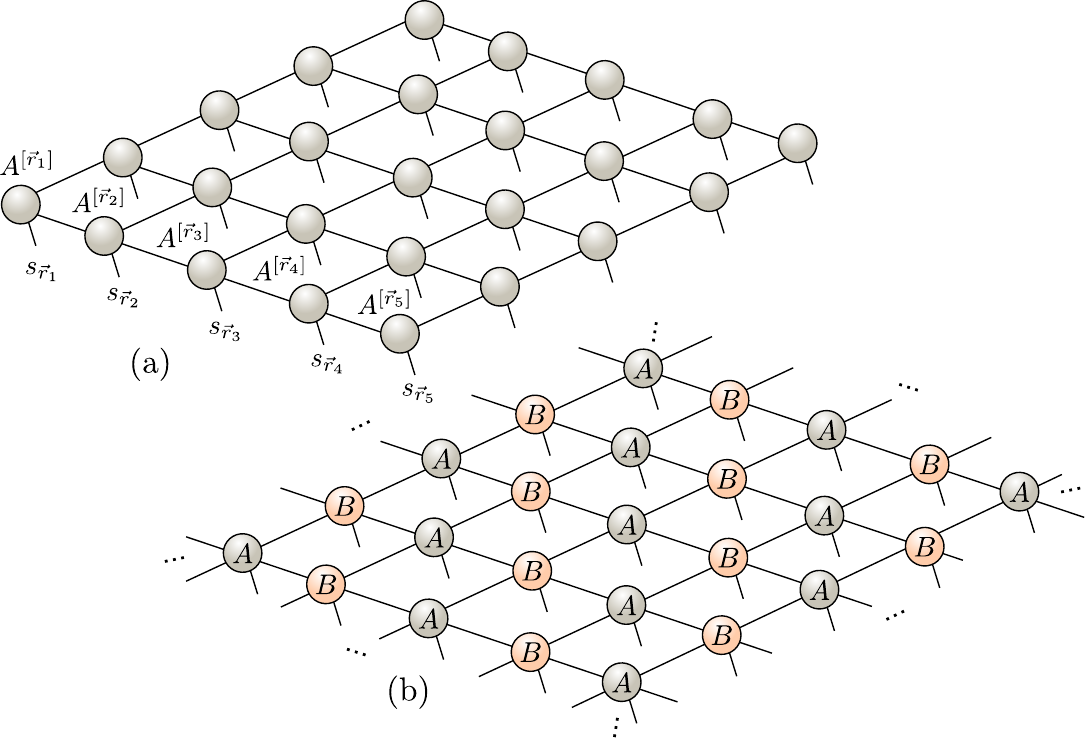}
	\caption{(Color online) (a) Graphical representation of a PEPS on a $5\times 5$ square lattice. Each ball is a tensor, and lines correspond to tensor indices. Lines from one tensor to another correspond to common summed indices (or \emph{contracted} indices). The free index, or open leg, in each tensor is called \emph{physical index}, and corresponds to the local degrees of freedom of the local Hilbert space at every site. (b) An infinite PEPS with a two-tensor unit cell. The two tensors $A$ and $B$ are repeated on the infinite $2d$ lattice.}
	\label{PEPS}
\end{figure}

PEPS have some important properties that make it an appropriate representation for 2d quantum lattice systems. First, and similarly to MPS, PEPS satisfy the entanglement area law.~\cite{arealaw1,arealaw2,arealaw3,arealaw4,arealaw5} More specifically, the scaling of entanglement entropy of an $L\times L$ block of a PEPS is $\mathcal{O}(L\log D)$. Accordingly, PEPS can describe very well the entanglement structure of many interesting 2d quantum systems, including low-energy eigenstates of many 2d Hamiltonians with local interactions. Second, PEPS can in principle represent systems with both \emph{finite} and \emph{infinite} correlation length.~\cite{Frank_corr} This is to be contrasted with the case of MPS, where only a finite correlation length is possible. Third, and also unlike for MPS, given the loops present on 2d lattices there is no obvious canonical form for a PEPS (although some proposals have been recently put forward along this direction \cite{Phiencano1,Romancano1,Romancano2}). Last, but not least, PEPS can be used to represent systems in the \emph{thermodynamic limit} by using a small number of tensors, under the assumption of shift invariance. More precisely, a unit cell of tensors is repeated all over the 2d lattice to construct an arbitrarily-large shift-invariant PEPS. For an infinite system, this is the so-called infinite-PEPS, or iPEPS (see Fig.~\ref{PEPS}(b) for a graphical illustration of an iPEPS with a two-site unit cell).

\subsubsection{Numerical application}

PEPS can be used to study both ground state properties as well as dynamics of 2d quantum lattice systems. For ground states, one can either (i) variationally optimize the PEPS tensors so as to minimise the expectation value of the 2d corresponding Hamiltonian (as done in DMRG in 1d), or (ii) evolve the system in imaginary time until a fixed point (ground state) is reached (as done in TEBD in 1d). The second method can also be applied to study real time evolution of the system. This approach is also easy to extend to the thermodynamic limit, called the iPEPS algorithm,~\cite{Jordan1,Roman2} which we go over again briefly in the next section.

\subsection{\label{secIII} The iPEPS algorithm}
\subsubsection{Generalities}
For a given Hamiltonian $H$, the ground state of the system can be obtained by evolving an initial state $\ket{\Psi_0}$ in imaginary time $\beta$ as described by
\bea
\ket{\Psi_{\GS}} &=& \lim_{\beta \rightarrow \infty} \frac{e^{-H\beta}\ket{\Psi_0}}{||e^{-H\beta}\ket{\Psi_0}||}.
\label{TEBD1}
\eea
Moreover, the real-time evolution is described via the solution of the Schr\"odinger equation, which for a time-independent Hamiltonian $H$ reads
\bea
\ket{\Psi(t)} &=&e^{-iHt}\ket{\Psi_0}.
\label{ReTEBD1}
\eea
From now on, let us consider the evolution of an iPEPS in imaginary time (the extension to real-time evolution is technically straightforward). 

We assume that the Hamiltonian contains only translationally invariant nearest-neighbour interactions, i.e.,
\bea
H = \sum_{\langle \vec{r},\vec{r'} \rangle}h^{[\vec{r},\vec{r}']},
\eea
where the sum is performed over all the nearest neighbours $ \langle \vec{r},\vec{r}' \rangle$. For a Hamiltonian that is invariant under translations, we can use an iPEPS also with translation symmetry to represent the wave function $\ket{\Psi}$ of the system. This could be achieved by repeating the same tensor on each lattice site all over the lattice. However, our update scheme (described below) requires neighboring tensors to be different, which is why we typically use a two-site translationally invariant iPEPS shown in Fig.~\ref{PEPS}(b), depending only on two tensors $A$ and $B$. If translational symmetries are spontaneously broken in the thermodynamic limit, then a larger unit cell of tensors (compatible with the structure of the ground state) can be used instead.~\cite{Philippe3}


In order to compute the ground state of the system by an evolution in imaginary time, one first decomposes the time evolution operator into a product of so-called \emph{two-body gates}. To this end, the Hamiltonian $H$ is rewritten as
\bea
H = H_{l}+H_{r}+H_{u}+H_{d},
\eea
where each term  $H_{i} = \sum_{\langle \vec{r},\vec{r}' \rangle \in{i}}h^{[\vec{r},\vec{r}']}$, $i\in(l,r,u,d)$ is the sum of mutually-commuting Hamiltonian terms for links labelled as (\emph{left, right, up, down}). Notice, though, that  the commutator $[H_{i},H_{j}]$ is in general different from zero whenever $i \neq j$. Applying the first-order Suzuki-Trotter decomposition,~\cite{Suzuki1} we can write the time evolution operator as 
\bea
e^{-H\beta} &=& (e^{-H\delta})^{m}\nn\\
&\approx&(e^{-H_l\delta}e^{-H_r\delta}e^{-H_u\delta}e^{-H_d\delta})^{m}
\label{suzu1}
\eea
where $\delta$ is an infinitesimal time step and $m \equiv \beta/\delta\in\mathbb{N}$ is the number of "time steps" that needs to evolve the system, so as to reach total evolution time $\beta$. Since each term $H_{i}$ is a sum of mutually-commuting  terms, we can further write each term $e^{-H_{i}\delta}$ in Eq.~(\ref{suzu1}) as a product of two-body gates, i.e., 
\bea
e^{-H_{i}\delta}&=&\prod_{\langle \vec{r},\vec{r'} \rangle \in{i}}g^{[\vec{r},\vec{r}']}, 
\label{suzu2}
\eea
with $g^{[\vec{r},\vec{r}']} \equiv e^{-h^{[\vec{r},\vec{r}']}\delta}$. 

For a given time step of the evolution, let us consider the action of the term $e^{-H_{i}\delta}$ on an infinite PEPS $\ket{\Psi_{AB}}$ with two tensors $A$ and $B$, and bond dimension $D$. To this aim we focus on \emph{one} link on the lattice and, \emph{only initially}, disregard the effect of the gates $g^{[\vec{r},\vec{r}']}$ on the rest of the links (which is approximately correct, since $\delta \ll 1$ and thus $g^{[\vec{r},\vec{r}']} \sim {\mathbb I}$). Let us consider the case of an $r-$link, for concreteness. After applying the gate, we obtain a new iPEPS $\ket{\Psi_{A'B'}}$ which is characterized by tensors $A$ and $B$ everywhere except for the two tensors connected by the link where the gate acted. More precisely, $\ket{\Psi_{A'B'}}$ and $\ket{\Psi_{AB}}$ differ from each other only by two tensors. Because of the effect of the gate, the bond dimension of the affected index changes  to $D '\leq d^{2} D$, and thus increases, corresponding to a change of the entanglement in the tensor network. Because of this, the iPEPS bond dimension quickly increases exponentially fast after few gate applications, making the simulation intractable. To overcome this problem, the infinite PEPS $\ket{\Psi_{A'B'}}$ is approximated by a new PEPS $\ket{\Psi_{\tilde{A}\tilde{B}}}$, by replacing tensors $A'$ and $B'$ by $\tilde{A}$ and $\tilde{B}$, where these two last tensors have again bond dimension $D$ for the affected index. This is done in a way such that the state $\ket{\Psi_{\tilde{A}\tilde{B}}}$ is close to the exact state $\ket{\Psi_{A'B'}}$, and thus introduces a small error only. Such a procedure is called \emph{tensor update} of the iPEPS.  

A possibility to implement the tensor update is to look for new tensors $\tilde{A}$ and $\tilde{B}$ that minimise the squared distance between the exact and the approximating state, i.e.
\bea
 \min_{\tilde{A},\tilde{B}}|| \ket{\Psi_{A'B'}}-\ket{\Psi_{\tilde{A}\tilde{B}}}||^{2} =  \min_{\tilde{A},\tilde{B}} d(\tilde{A},\tilde{B}), 
 \label{figureofmerit}
\eea
with 
\bea
d(\tilde{A},\tilde{B}) &=& \braket{\Psi_{A'B'}}{\Psi_{A'B'}}+\braket{\Psi_{\tilde{A}\tilde{B}}}{\Psi_{\tilde{A}\tilde{B}}}-\nn\\
&&-\braket{\Psi_{\tilde{A}\tilde{B}}}{\Psi_{A'B'}}-\braket{\Psi_{\tilde{A}\tilde{B}}}{\Psi_{A'B'}}.
\label{costfunction}
\eea
To solve the problem in Eq.~(\ref{figureofmerit}), two main tasks are needed. These are \emph{(i) the effective environment calculation}, and \emph{(ii) the tensor update}. Since these pieces of the method will be fundamental for the improvements to be later explained, we review them in detail in the following.

\subsubsection{Effective environment calculation} 

To properly evaluate $d(\tilde{A},\tilde{B})$ one needs to take into account the effect of the whole tensor network surrounding the affected link, i.e., the \emph{environment}. Such a tensor network of infinitely-many tensors is conveniently approximated by an \emph{effective environment}, consisting of a small number of tensors only. The effective environment can be computed using various approaches, such as TRG/SRG, HOTRG/HOSRG,~\cite{Levin1,Xie1,Zhao1,Xie2} tensor network renormalization (TNR),~\citep{GlenRenormT} iTEBD,~\cite{Jordan1} and corner transfer matrix (CTM) methods. This last approach will be our choice in this paper. We shall not explain CTM methods in full detail here, and we refer the reader to the extensive existing literature on the topic such as Refs.~[\onlinecite{Baxter11,Nishino1,Nishino1_1,Roman2,Philippe4,RomanNewCTM1}] for technicalities. However, since these methods will also be important at a later stage of this paper, we review briefly some notations and conventions. 

\begin{figure}
	\includegraphics[scale=0.75]{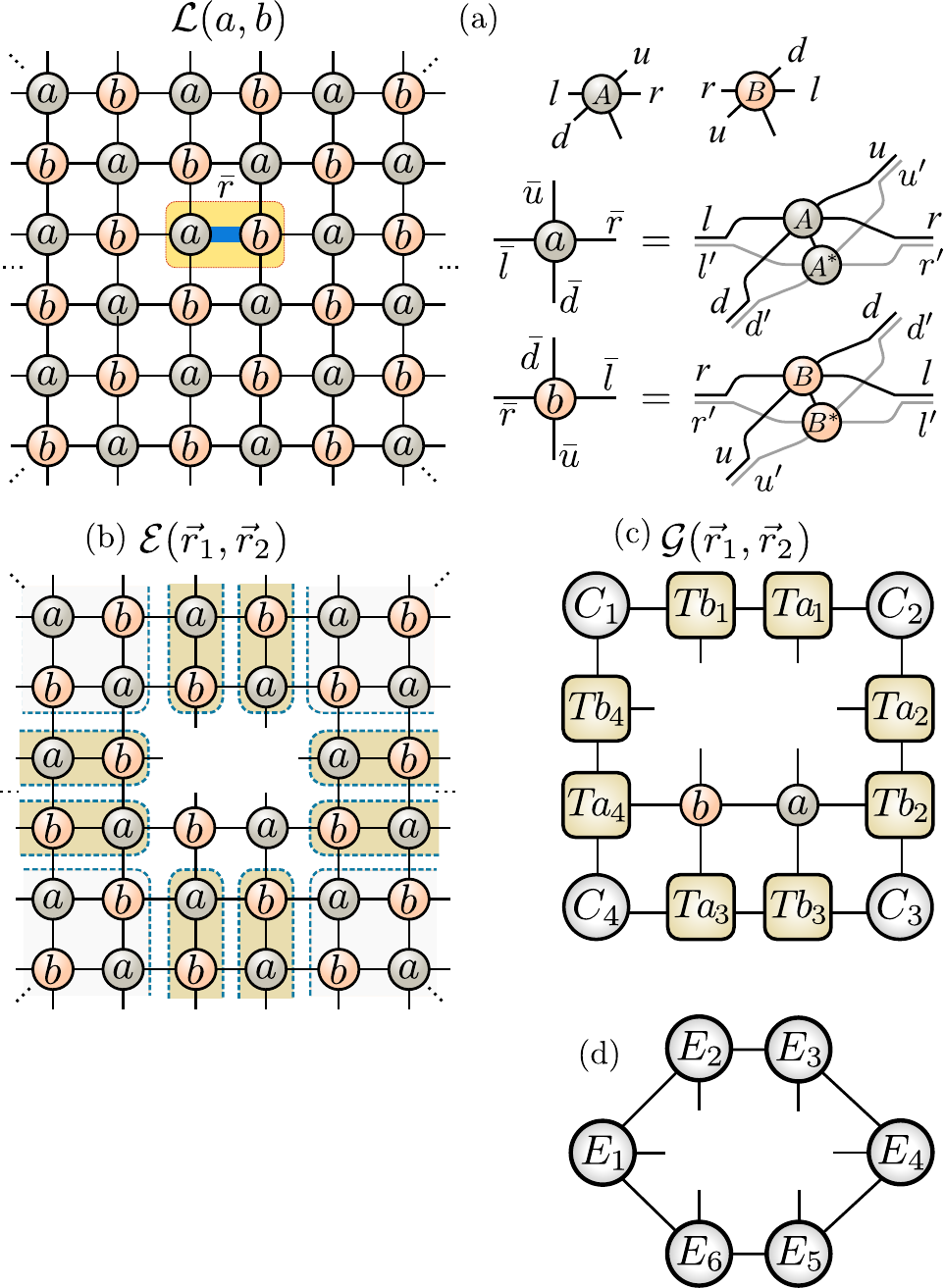}
	\caption{(Color online) (a) Left: 2d lattice of tensors formed from $a$ and $b$; right: contractions to obtain tensors $a$ and $b$. (b) Environment of a given link on the lattice (here an r-link). (c) Effective environment of a given link on the lattice (here an r-link). (d) 6-tensor representation of the effective environment around the link.}
    \label{CTM_Env}
\end{figure}

We first construct an infinite square lattice $\mathcal{L}(a,b)$ by contracting the physical indexes of  $\ket{\Psi_{AB}}$ and $\bra{\Psi_{AB}}$, see Fig.~\ref{CTM_Env}(a) where 
\bea
a_{\bar{l}\bar{r}\bar{u}\bar{d}} &=& \sum_{s}A^{s}_{lrud}(A^{s}_{l'r'u'd'})^{*}\\
b_{\bar{l}\bar{r}\bar{u}\bar{d}} &=& \sum_{s}B^{s}_{lrud}(B^{s}_{l'r'u'd'})^{*},
\eea
where $\bar{l}$, $\bar{r}$, $\bar{u}$, $\bar{d}$ are combined indices, e.g. $\bar{l}=(l, l')$.

The exact environment $\mathcal{E}(\vec{r}_1,\vec{r}_2)$ of two sites at $\vec{r}_1$ and $\vec{r}_2$ is shown in Fig.~\ref{CTM_Env}(b). In CTM methods, this is approximated by an effective tensor network $\mathcal{G}(\vec{r}_1,\vec{r}_2)$, the effective environment, which comprises a set of four $\chi \times \chi$ corner transfer matrices $\{C_{1}, C_{2}, C_{3}, C_{4}\}$, eight half transfer row/column tensors $\{Ta_{1}, Ta_{2}, Ta_{3}, Ta_{4}, Tb_{1}, Tb_{2}, Tb_{3}, Tb_{4}\}$ and two tensors $a$ and $b$, see Fig.~\ref{CTM_Env}(c). A further simple contraction of the tensor network produces an effective environment for the considered link in terms of only six tensors $\{E_{1}, E_{2}, E_{3}, E_{4}, E_{5}, E_{6}\}$, see Fig.~\ref{CTM_Env}(d).

\subsubsection{iPEPS Tensor Full Update} 

\begin{figure}
	\includegraphics[width=\columnwidth]{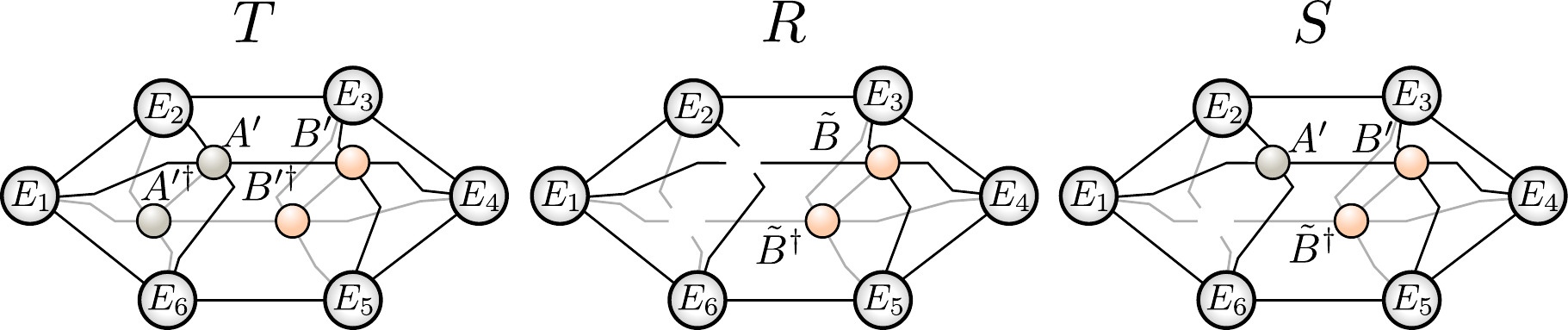}
	\caption{(Color online) Tensor network diagrams showing how to compute $T, R$ and $S$ in Eq.~(\ref{var1}).}
	\label{updateTensors1}
\end{figure}

Eq.~(\ref{costfunction}) implies that the squared distance $d(\tilde{A},\tilde{B})$ is a quadratic function of tensors $\tilde{A}$ and $\tilde{B}$. Thus, once the effective environment is obtained, one can apply an alternating least squares method to optimize tensors $\tilde{A}$ and $\tilde{B}$. This is the so-called \emph{full update} (FU), and it follows the steps below: 

\begin{itemize}
 \item[(i)] Fix tensor $\tilde{B}$ to some initial tensor\footnote{A good initialization of $\tilde{A}$ and $\tilde{B}$ can be obtained by applying the following procedure. First, contract the two tensors $A$ and $B$ along the link of interest together with the two-body gate $g$. Second, apply a singular value decomposition of the obtained tensor separating the links associated to each lattice site. Finally, truncate the common link up to $D$ values, leads to a good candidate for $\tilde{A}$ and $\tilde{B}$.} or to the tensor obtained from previous iteration. In order to find $\tilde{A}$, rewrite  Eq.~(\ref{costfunction}) as a quadratic scalar expression for the tensor, 
\bea
d(\tilde{A},\tilde{A}^{\dagger})&=&\tilde{A}^{\dagger}R\tilde{A}-\tilde{A}^{\dagger}S-S^{\dagger}\tilde{A}+T.
\label{var1}
\eea
In this equation, $\tilde{A}$ is understood as a reshaped vector with $D^{4}d$ components and matrices $R, S, T$ can be obtained from the appropriate tensor contractions including the effective environment around the $r$-link, see Fig.~\ref{updateTensors1}.

\item[(ii)] We find the minimum of $d(\tilde{A},\tilde{A}^{\dagger})$ in Eq.~(\ref{var1}), with respect to $\tilde{A}^{\dagger}$, which is given by $\tilde{A} =R^{-1} S$. 

\item[(iii)] Next, we fix the tensor $\tilde{A}$ and find $\tilde{B}$ using the same procedure as steps (i) and (ii) above.
\end{itemize}

The above steps are iterated until the cost function $d(\tilde{A},\tilde{B})$ converges to a sufficiently-small value. A possibility to check convergence is, e.g., to check the value of this cost function between two successive iterations and compare it to some small tolerance. Last but not least: once the optimal tensors are found, \emph{these are replaced over the entire 2d lattice} approximating the effect of all the gates $g^{[\vec{r},\vec{r}']}$ acting over the infinitely-many links of the same type. Such a procedure defines the updated infinite PEPS $\ket{\Psi_{\tilde{A}\tilde{B}}}$ in terms of the two new tensors. Finally, the same procedure is repeated for the $l$-, $u$- and $d$-links to complete one time step. This is iterated until the desired real running time is achieved, or until the iPEPS converges to a fixed-point approximation of the ground state for imaginary-time evolutions. 

As such, the computational cost of the procedure above is quite high due to the calculation of the inverse of the matrix $R$ in step (ii). Specifically, this cost is $\mathcal{O}({D}^{8})$ if recurrent methods are applied for computing the inverse (e.g., biconjugate gradient). Otherwise, the cost would be $\mathcal{O}({D}^{12})$ for an exact inverse calculation.

\subsubsection{Reduced Tensor Full Update}Ê
\label{sec:reducedupdate}
As an efficient and convenient alternative to this method, one can apply the "revised" FU scheme discussed in Ref.~\onlinecite{Philippe1} where, instead of updating two tensors  $\tilde{A} $ and $\tilde{B}$, one updates some lower-rank \emph{subtensors} related to them (sometimes they are also called \emph{reduced tensors}\cite{Lubasch2}). These subtensors are denoted as $\tilde{a}_R$ and $\tilde{b}_L$, respectively. The main idea of this optimization scheme is based on the observation that applying the two-body gate $g$ on two sites  $A $ and $B$ connected by a specific link changes the properties of that link \emph{only},  while the others remain unchanged. For instance, for an $r$-link this results into modifying the bond dimension for this link from ${D}$ to ${D}'>{D}$, while the size of the other indices is unaffected. 

\begin{figure}
	\includegraphics[scale = 0.8]{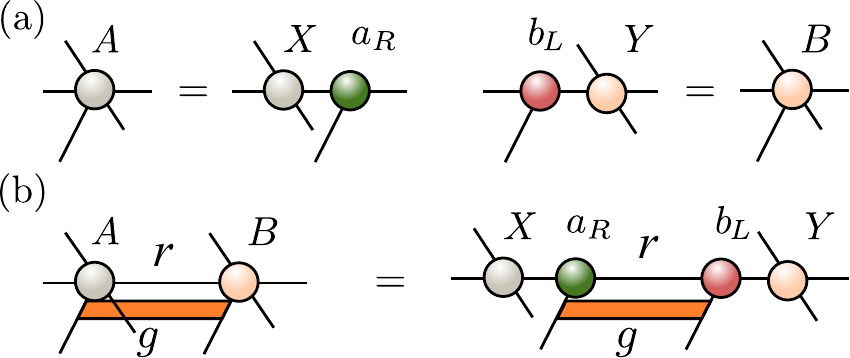}
	\caption{(Color online) (a) Subtensors for iPEPS tensor $A$ and $B$. Decompose $A = Xa_R$ and $B=b_LY$ by means of decompositions such as QR or SVD. (b) The action of the two-body gate $g$ on the iPEPS tensors $A$ and $B$ connected by an $r$-link is equivalent to its action on the subtensors $a_R$ and $b_L$ only, leaving $X$ and $Y$ unaffected.}
	\label{reducedTensors}
\end{figure}

\begin{figure}
	\includegraphics[width=\columnwidth]{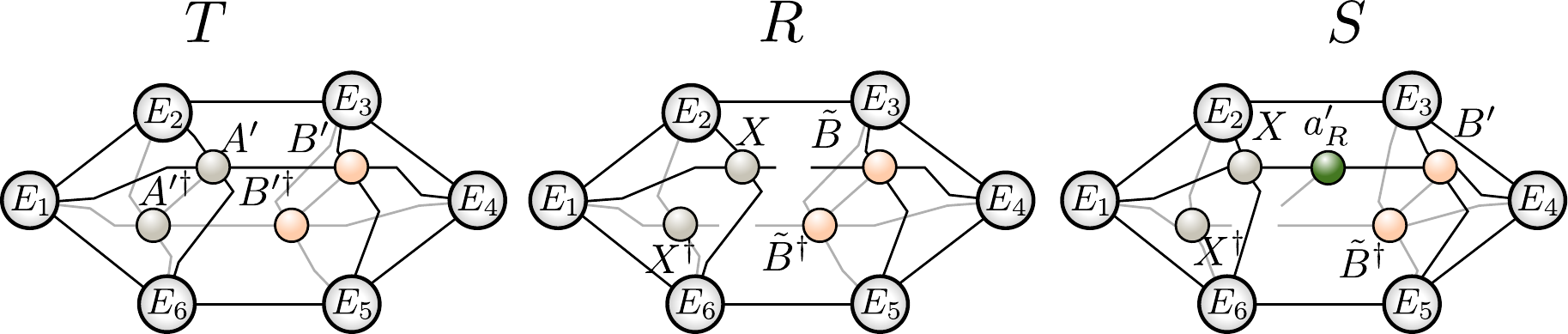}
		\caption{(Color online) Tensor network diagrams showing how to compute $T, R$ and $S$ in Eq.~(\ref{varreducesdtensor1}).}
	\label{updateTensors2}
\end{figure}

Thus, by means of the QR or singular value decomposition (SVD), we decompose tensors $A$ and $B$ such that $A= Xa_R$ and $B=b_LY$, where $a_R$ and $b_L$ are connected by the bond index on the $r$-link, and also contain the physical indices (see Fig.~\ref{reducedTensors}(a)). When applying the two-body gate $g$, we now contract it with $a_R$ and $b_L$ in order to update them directly. Once $a_R$ and $b_L$ are updated to, say, $\tilde{a}_R$ and $\tilde{b}_L$, we can easily get updated iPEPS tensors $\tilde{A}$ and $\tilde{B}$ by using $\tilde{A} = X\tilde{a}_R$ and $\tilde{B}=\tilde{b}_LY$. The update scheme for $\tilde{a}_R$ and $\tilde{b}_L$ can be performed similarly as we explained for $\tilde{A}$ and $\tilde{B}$, following steps (i-iii) above. More precisely, in these steps we replace $\tilde{A}$ and $\tilde{B}$ by $\tilde{a}_R$ and $\tilde{b}_L$, which are alternatively optimized according to a similar figure of merit defined as
\bea
d(\tilde{a}_R,\tilde{b}_L) &=& \braket{\Psi_{a'_Rb'_L}}{\Psi_{a'_Rb'_L}}+\braket{\Psi_{\tilde{a}_R\tilde{b}_L}}{\Psi_{\tilde{a}_R\tilde{b}_R}}-\nn\\
&&-\braket{\Psi_{\tilde{a}_R\tilde{b}_L}}{\Psi_{a'_Rb'_L}}-\braket{\Psi_{\tilde{a}_R\tilde{b}_L}}{\Psi_{a'_Rb'_L}}. 
\label{costfunction_reducedtensors}
\eea
Eq.~(\ref{var1}) for $\tilde{a}_R$ is now rewritten as,
\bea
d(\tilde{a}_R,\tilde{a}^{\dagger}_R)&=&\tilde{a}^{\dagger}_RR\tilde{a}_R-\tilde{a}^{\dagger}_RS-S^{\dagger}\tilde{a}_R+T,
\label{varreducesdtensor1}
\eea
and the cost function for variable $\tilde{b}_L$ is defined in a similar way. Note that the tensors $\{R,S,T\}$ in Eq.~\ref{varreducesdtensor1} are different from the ones in Eq.~\ref{var1}, see Fig.~\ref{updateTensors2}. 

The computational cost of this update scheme is $\mathcal{O}(d^3D^{6})$, where the inverse of the matrix $R$ can now be  computed \emph{exactly}. Due to this huge advantage in computational cost, as compared to the direct update of the iPEPS tensors $A$ and $B$, the present scheme is able to deal with iPEPS of larger bond dimension $D$, with the corresponding advantages. 

The rest of the update follows as explained in the previous subsection. This is, the reduced tensors are optimized until the cost function is sufficiently small, and then, these are replaced over the entire 2d lattice. As explained before, this last step approximates the effect of all the gates acting on all the links of the lattice of the same type. As before, the procedure defines an updated infinite PEPS $\ket{\Psi_{\tilde{A}\tilde{B}}}$ in terms of two new tensors. After this, one repeats the same procedure for the $l$-, $u$- and $d$-links to complete one time step of the evolution. This will be iterated until the iPEPS converges to a fixed point approximating the ground state (for imaginary time), or until the desired real-time evolution is completed. 

\section{Improving the iPEPS algorithm}


One of the main limitations of the FU iPEPS algorithm in its present formulation is that the effective environment has to be recomputed from scratch at every time step. Obtaining a converged effective environment requires $N_{\text{CTM}}$ iterations of the CTM algorithm with $N_{\text{CTM}}$ depending on the amount of entanglement in the system (for not-too-entangled systems, typically $N_{\text{CTM}} \sim 10 - 20$, but it can be considerably larger in strongly entangled systems). Since these CTM iterations are the computational bottleneck of the method, it is desirable to keep $N_{\text{CTM}}$ as low as possible. However, if one uses a $N_{\text{CTM}}$ which is too small, then the errors introduced in the effective environment can lead to instability problems of the method.

Here we propose two improvements to the algorithm, which enhance the stability of the method and  make it  more efficient. 
First, we explain how to implement what we call a \emph{fast full update} (FFU), where the accuracy of the FU is preserved while substantially reducing  its computational cost and improving its stability. Second, we discuss the application of a \emph{local gauge fixing}, as in Ref.~\onlinecite{Lubasch2}, which naturally improves stability and also accelerates the convergence of the overall method. This is explained in what follows, and benchmarking numerical calculations shall be provided in the forthcoming sections.

\subsection{Fast Full Update (FFU)}
In this update scheme we reduce the computational cost by using the following idea: instead of recomputing the environment tensors from scratch we can update the environment tensors \emph{simultaneously} with those of the iPEPS at each step, just by \emph{a single CTM iteration step}. This reduces the computational cost by a large factor (see results below). The crucial technical point of this FFU scheme is to make sure that the environment tensors remain compatible with the updated iPEPS tensors throughout the (imaginary) time evolution. 


The details of the FFU scheme for iPEPS tensors $A$ and $B$ plus the effective environment tensors are shown in Fig.~\ref{updatescheme}. In the following we explain 
 how to proceed at each time step, when the two-body gates $g$ are successively applied on the $r$-, $l$-, $u$- and $d$-links: 

\begin{itemize}
	\item[(a)] \underline{\it $r$-link update:} Suppose that the effective environment around four links of the iPEPS is characterized by tensors $\{C_i,Ta_i,Tb_i\}_{i=1}^{4}$, see Fig.~\ref{updatescheme}(a-i). Applying the gate $g$ to the $r$-link will modify the properties of this link, so we need to update the tensors $A$ and $B$ using the update scheme from the previous section. Specifically, the effective environment of an $r$-link is obtained by first absorbing a row of tensors $\{Ta_4, b, a, Tb_2\}$ to the bottom edge and then contracting the tensors appropriately, such that the effective environment is represented by six tensors $\{E_i\}_{i=1}^{6}$ as in Fig.~\ref{updatescheme}(a-ii). We then apply the tensor update scheme explained previously in order to find some updated tensors $A_1$ and $B_1$. 
	
	In order to prepare for the next update of the $l-$link, we now update the environment tensors. To this end, we insert two columns of tensors $\{Ta_1, b_1, a_1, Tb_3\}$ and $\{Tb_1, a_1, b_1, Ta_3\}$ in the middle of the tensor network, see Fig.~\ref{updatescheme}(a-iii,iv). Note that \emph{the two tensors $a$ and $b$ connected by link $\bar{r}$ are now replaced by the updated ones $a_1$ and $b_1$}. But importantly, \emph{there are still two sites where we use the old tensors $a$ and $b$}, since the corresponding $r$-link is connected to the old environment, and thus was not "formally" updated.
	 
Next, the columns of tensors $\{Ta_1, b_1, a, Tb_3\}$ and $\{Tb_1, a_1, b, Ta_3\}$ are absorbed to the right and left edges respectively, see Fig.~\ref{updatescheme}(a-iv). After this absorption, one finds appropriate isometries to renormalize the environment tensors as usual, in such a way that their bond dimensions do not grow.~\cite{Roman2} The new environment tensors of the $l$-link are denoted as in Fig.~\ref{updatescheme}(b-i).
	\begin{figure}
		\includegraphics[width=\columnwidth]{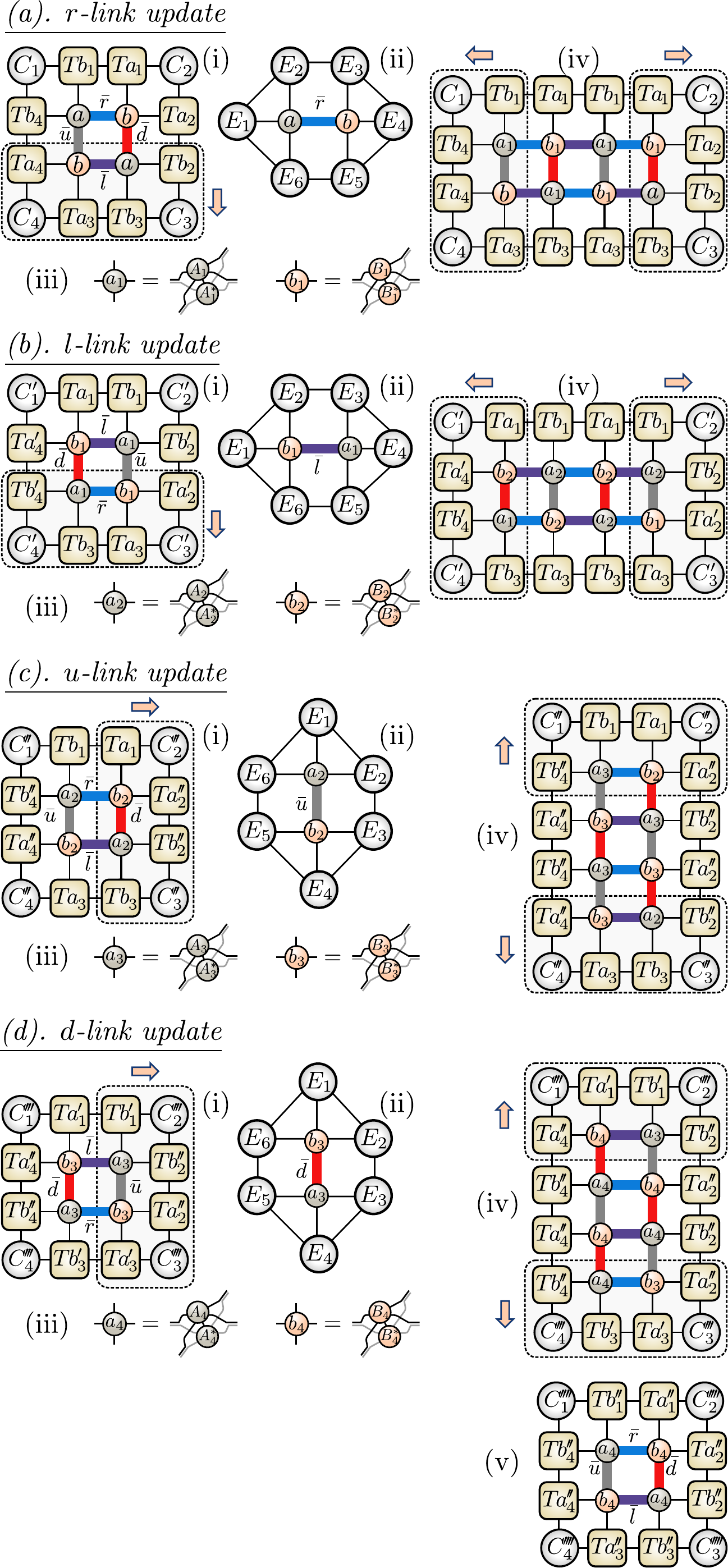}
		\caption{\label{updatescheme} (Color online) Tensor network diagrams for the Fast Full Update. Details are explained in the main text.}
	\end{figure}
	\item[(b)] \underline{\it $l$-link update:} For the action of a gate on the $l$-link, we proceed in the same way as with the $r$-link, which is shown graphically in Fig.~\ref{updatescheme}(b). The iPEPS is now updated and represented by two new tensors $A_2$ and $B_2$. The new environment tensors for updating the $u$-link are denoted as in the Fig.~\ref{updatescheme}(c-i). 
	\item[(c)] \underline{\it $u$-link update:} From the tensor network in Fig.~\ref{updatescheme}(c-i), we obtain the environment tensors for the $u$-link as shown in Fig.~\ref{updatescheme}(c-ii). An update of the tensors will now produce two new tensors $A_3$ and $B_3$. We then compute the effective environment for the update of the $d$-link as shown in Fig.~\ref{updatescheme}(c-iv)
	\item[(d)] \underline{\it $d$-link update:} Now we follow the same procedure as for the $u$-link in order to update the iPEPS  tensors as well as the effective environment. This is shown in Fig.~\ref{updatescheme}(d). Finally, we obtain the new iPEPS tensors $A_4$ and $B_4$ and the new effective environment of the four links is represented by tensors $\{C''''_i,Ta''_i,Tb''_i\}_{i=1}^{4}$, see Fig.~\ref{updatescheme}(d-v).
\end{itemize}

The above update scheme is illustrated for one time step only. One then needs to iterate this scheme in order to evolve the system up to the desired time. As in the update approaches explained in the previous section, for the FFU is also a good idea to choose properly the initial state in ground-state calculations (e.g. converged iPEPS obtained from the SU scheme, or from the FU scheme with a smaller bond dimension). Evidently, for the FFU this choice also helps in the stability and fast convergence of the algorithm. 

The FFU that we just presented has two key advantages: first, \emph{one keeps an environment at every step that is perfectly compatible with the tensors in the iPEPS in all bond indices}. This is the reason why, e.g., in Fig.~\ref{updatescheme}(a-iv) we still have two tensors $a$ and $b$ at two sites. Such a property naturally improves the stability. Second, the environment \emph{is not reconverged for every link at every step}.\color{red}Ê\footnote{Formally speaking, this means that the FFU is not the "correct" update, in the sense that the environment tensors are not fully converged. In fact, it does not even correspond to the full first order correction in perturbation theory for the environment. Still, the point of this paper is to see that the FFU is already sufficient for  practical purposes.} \color{black} As a result, this reduces the required computational time considerably.  

\subsection{Gauge fixing}
In contrast to MPS, a PEPS does not have a canonical form. Therefore, it is difficult to fix the gauge of the PEPS tensors in some appropriate way during the time evolution. Despite this, local gauge fixing schemes have proven useful in improving the stability of the algorithm.~\cite{Phiencano1,Romancano1,Romancano2,Lubasch2} In this paper, we use the local gauge fixing proposed in Ref.~\onlinecite{Lubasch2} (applied there to finite PEPS), in the context of the iPEPS algorithm. We shall see that this not only helps to  improve the stability of the method, but also to accelerate its convergence. 

\begin{figure}
		\includegraphics[width=\columnwidth]{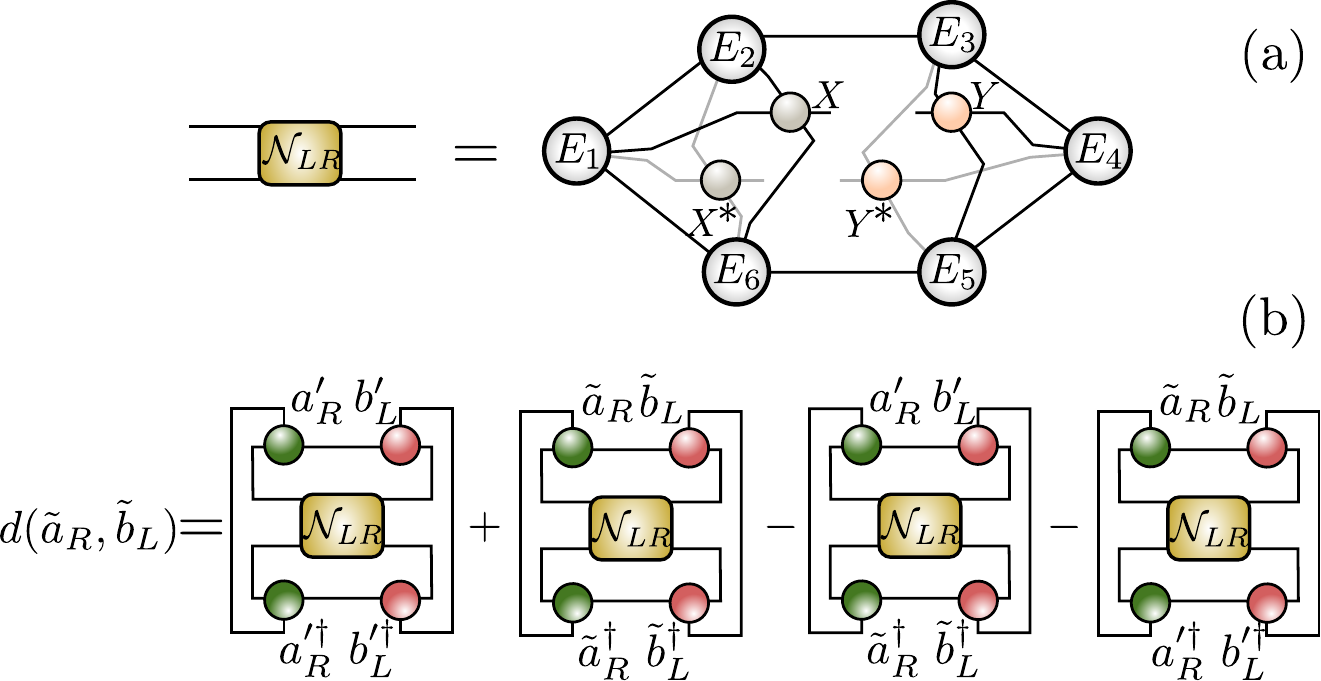}
		\caption{\label{CostFunction_reducedtensor} (Color online) (a) Contraction producing the norm tensor. The leading computational cost of this contraction is $\mathcal{O}(d^4D^4\chi^2)$ + $\mathcal{O}(d^2D^6\chi^2)$+$\mathcal{O}(d^2D^3\chi^3)$\cite{Lubasch2}.  (b) Diagrammatic representation of the cost function defined in Eq.~(\ref{costfunction_reducedtensors1}).}
	\end{figure}
	
To this end, we consider the tensor update applied to the reduced tensors. We rewrite Eq.~(\ref{costfunction_reducedtensors}) as follows: 
\bea
d(\tilde{a}_R,\tilde{b}_L) &=& a'^{\dagger}_Rb'^{\dagger}_L\mathcal{N}_{LR}a'_Rb'_L+\tilde{a}^{\dagger}_R\tilde{b}^{\dagger}_L\mathcal{N}_{LR}\tilde{a}_R\tilde{b}_R-\nn\\
&&-\tilde{a}^{\dagger}_R\tilde{b}^{\dagger}_L\mathcal{N}_{LR}a'_Rb'_L-\tilde{a}^{\dagger}_R\tilde{b}_L\mathcal{N}_{LR}a'_Rb'_L,
\label{costfunction_reducedtensors1}
\eea
where $\mathcal{N}_{LR}$ is the "norm" tensor obtained by computing  the overlap of $\braket{\Psi_{\tilde{a}_R\tilde{b}_L}}{\Psi_{\tilde{a}_R\tilde{b}_R}}$ while leaving out the tensors $\tilde{a}_R,\tilde{b}_L$ and their complex conjugates. Specifically, this can be done by contracting the tensor network shown in Fig.~\ref{CostFunction_reducedtensor} (a). Also, the cost function is represented diagrammatically in Fig.~\ref{CostFunction_reducedtensor} (b). In order to update the subtensors one needs, thus, to compute the norm tensor~$\mathcal{N}_{LR}$.

\begin{figure}
		\includegraphics[scale=0.75]{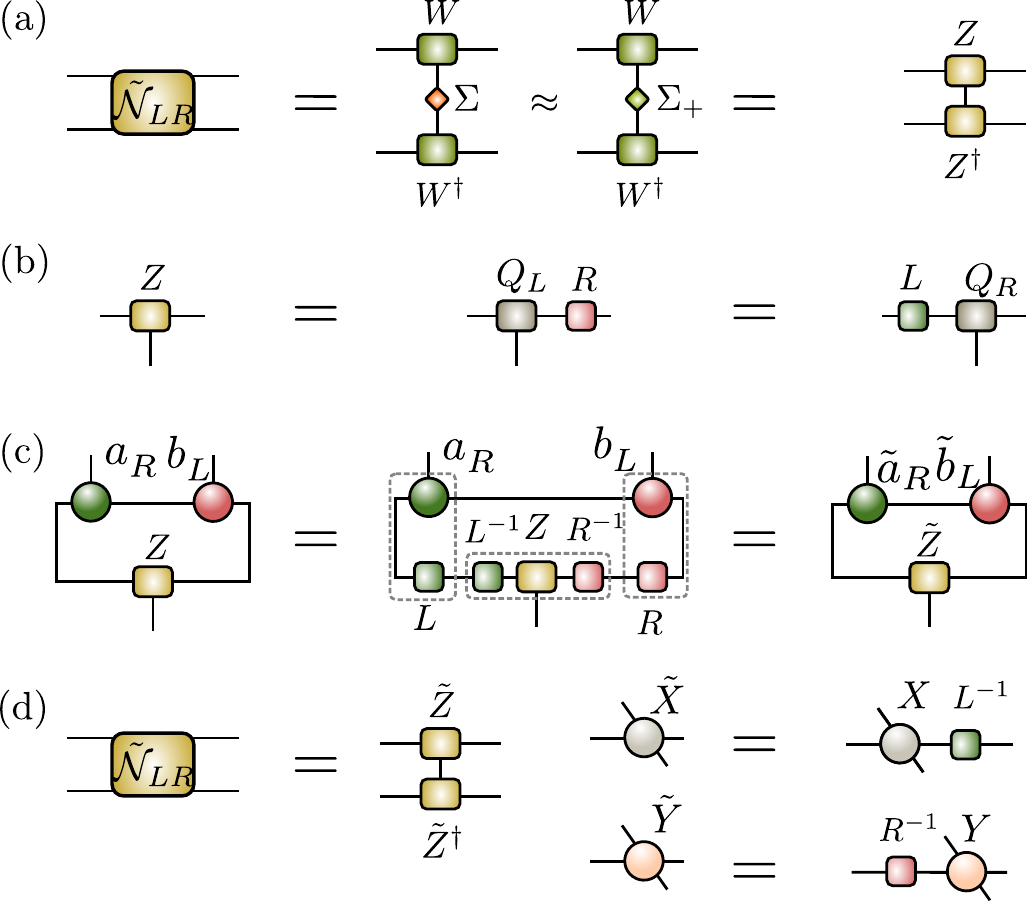}
		\caption{\label{Positive_GaugeNorm} (Color online) (a) Positive approximant for the norm tensor $\tilde{\mathcal{N}}_{LR}$.  (b) Apply the QR decomposition for the left and the right bond indices of tensor $Z$, so that $Z=Q_LR = LQ_R$. (c) Insert the identities $I = L^{-1}L = R^{-1}R$ into the left and right bond indexes of the tensor $Z$ to obtain $\tilde{Z} = L^{-1}ZR^{-1}$ and the new subtensors $\tilde{a}_{R} = La_R, \tilde{b}_{L} = b_LR$. (d) A new norm tensor is obtained as $\tilde{\mathcal{N}}_{LR} = \tilde{Z}\tilde{Z}^{\dagger}$. In order to keep the compatibility once the subtensors have been updated, we need to multiply  the tensors $X$ and $Y$ by matrices $L^{-1}, R^{-1}$, so that $\tilde{X} = XL^{-1}$ and $\tilde{Y} = R^{-1}Y$, respectively. This is done before recovering the tensors $\tilde{A} = \tilde{X}\tilde{a}_R$ and $\tilde{B} = \tilde{b}_L\tilde{Y}$.}
	\end{figure}

Tensor $\mathcal{N}_{LR}$ has the following properties: first, it is impossible to choose a gauge in such a way that this tensor is the identity matrix for all links at the same time. The reason is that an iPEPS does not have a canonical form in the same sense as MPS. Second, this tensor needs to be computed using approximations, which implies that, generally, it is neither strictly Hermitian nor positively defined. Although one could always apply some approximation methods that preserve positivity explicitly (such as the single layer method, \cite{Pizorn1}), the tensors obtained from such approaches do not produce as accurate results as other methods that do not enforce positivity (such as CTM methods).~\footnote{In fact, end even more fundamental, it has also been shown that some positive Matrix Product Operators with finite bond dimension do not have a representation in terms of a Matrix Product Density Operator with finite bond dimension, see Ref.~\onlinecite{MPOMPDO}.} It is well-known that the (normally small) negative part of the norm tensor often causes some ill-posed conditions in updating the iPEPS tensors. To circumvent this problem, one can  make it Hermitian and positive-defined using the following steps:~\cite{Lubasch2} first, one approximates the norm tensor as $\tilde{\mathcal{N}}_{LR} = (\mathcal{N}_{LR}+\mathcal{N}_{LR}^{\dagger})/2$, which is Hermitian. Next, we approximate $\tilde{\mathcal{N}}_{LR}$ by its positive part. To achieve this, one applies the eigenvalue decomposition $\tilde{\mathcal{N}}_{LR} = W\Sigma W^{\dagger}$, and replaces the (small) negative eigenvalues in $\Sigma$ by zero. This is the so-called \emph{positive approximant}. The approximate eigenvalues are now denoted $\Sigma_{+}$. Moreover, $\tilde{\mathcal{N}}_{LR} = ZZ^{\dagger}$, where $Z = W\Sigma_{+}^{1/2}$, see Fig.~\ref{Positive_GaugeNorm}(a).		

The gauge-fixing that we apply here is explained in Fig.~\ref{Positive_GaugeNorm}. After fixing the local gauge in the norm tensor, we replace $\tilde{\mathcal{N}}_{LR}$ and compatible subtensors into Eq.~(\ref{costfunction_reducedtensors1}), and then start the variational update of  subtensors for the iPEPS. As shown in Ref.~\onlinecite{Lubasch2}, this choice of gauge improves a lot the conditioning of the norm tensor, and thus greatly increases the stability of the  update. In particular, when this gauge-fixing is combined with the FFU, the resulting iPEPS algorithm is remarkably fast, stable and accurate.

\section{\label{secIV} Results}

%

\begin{table*}
	\centering	
	\begin{tabular}{l l l l l}
		$(D,\chi)$ & No Gauge & Gauge & $R$ & $L$\\	
		\hline\hline
		Heisenberg model&~&~\\
		\hline
		(2,20) & $(9.33\pm 1.50)\times 10^{2}$ & $1.46\pm 0.01$& $5.49\pm 0.23$& $5.29\pm 0.39$\\
		(3,30) & $(1.6\pm 0.17)\times 10^{6}$ & $(0.15\pm 0.00)\times 10^{2}$& $(0.41\pm 0.01)\times 10^{2}$& $(0.45\pm 0.01)\times 10^{2}$\\
		(4,40) & $(8.36\pm 0.30)\times 10^{4}$ & $(3.02\pm 0.02)\times 10^{1}$ & $(1.78\pm 0.00)\times 10^{1}$&$(1.80\pm 0.02)\times 10^{1}$\\
		(5,50) & $(3.46\pm 0.16)\times 10^{6}$ & $(1.04\pm 0.08)\times 10^{3}$& $(5.04\pm 0.06)\times 10^{1}$& $(5.03\pm 0.06)\times 10^{1}$\\
		(6,70) & $(1.01\pm 0.30)\times 10^{7}$ & $(2.21\pm 0.03)\times 10^{4}$& $(5.12\pm 0.15)\times 10^{1}$& $(5.05\pm 0.22)\times 10^{1}$\\
\hline
		Ising model &$h = 3.01$&~\\
		\hline
		(2,20) & $(3.44\pm 1.85)\times 10^{4}$ & $5.97\pm 0.09$ & $(1.29\pm 0.13)\times 10^{1}$ & $(1.29\pm 0.14)\times 10^{1}$\\
		(3,30) & $(4.28\pm 0.59)\times 10^{6}$ & $(1.95\pm 0.24)\times 10^{3}$ & $(5.00\pm 0.34)\times 10^{1}$& $(5.01\pm 0.32)\times 10^{1}$\\
		(4,30) & $(1.28\pm 1.52)\times 10^{19}$ & $(5.81\pm 11.88)\times 10^{17}$ & $(5.39\pm 0.13)\times 10^{2}$ & $(5.47\pm 0.14)\times 10^{2}$\\
			\end{tabular}	
	\caption{Mean condition number of the norm matrix $\tilde{\mathcal{N}}_{LR}$ as well as of matrices $L$ and $R$ with their standard deviation. The numbers are for the case of positive approximant, fixing and without fixing the gauge, for the subtensor FFU. We show different bond dimensions $(D,\chi)$ for an iPEPS algorithm with time step $\delta = 0.02$. The mean is computed over ten time steps, for Heisenberg and Ising models. The transverse field for the Ising model is $h = 3.01$, close to criticality.}
	\label{tab1}
\end{table*}

\begin{figure}
	\includegraphics[width =\columnwidth]{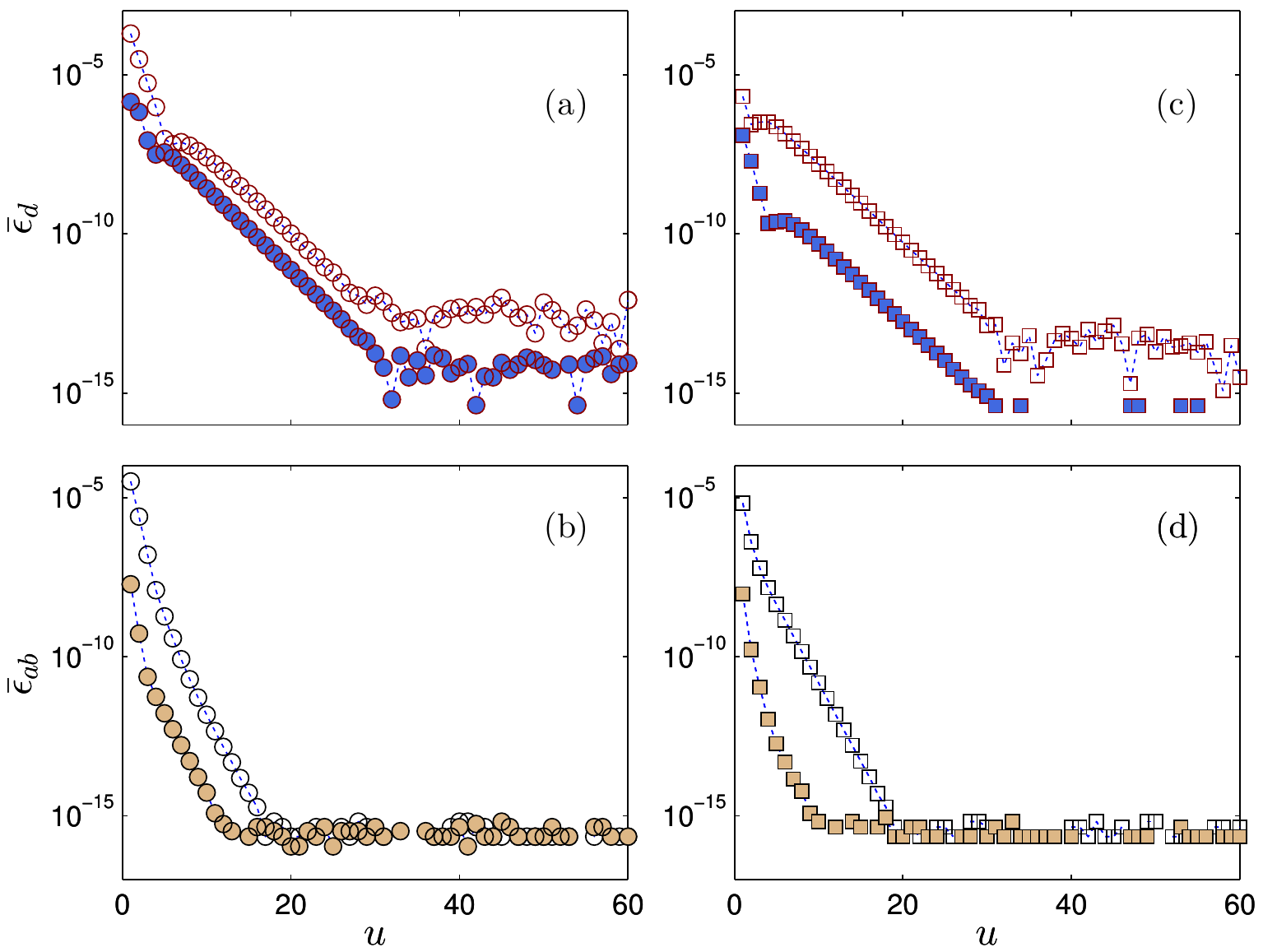}
	\caption{\label{costfunction} (Color online) Mean value of the relative change $\bar{\epsilon}_{d} = |d_{u+1}-d_{u}|/|d_{\text{init}}|$ of the cost function $d$ and the local fidelity $\bar{\epsilon}_{ab} = |f_{ab}^{u+1}-f_{ab}^{u}|/|f_{ab}^{\text{init}}|$ in the iPEPS algorithm with gauge (filled symbols) or without gauge (open symbols), for the Ising model with transverse field (a,b) and the Heisenberg model (c,d) respectively. For the Ising model the magnetic field is $h = 3.01$ and $(D,\chi) = (3,30)$, whereas for the Heisenberg model $(D,\chi) = (5,50)$. The time step is $\delta = 0.005$. The mean value is taken over all steps in imaginary time. 
	}
\end{figure}

We have benchmarked the improved iPEPS algorithm, including both the FFU and the gauge-fixing, by studying ground state properties of two models on an infinite 2d square lattice. The first model is the ferromagnetic quantum Ising model with transverse magnetic field, 
\bea
H = -\sum_{\langle \vec{r},\vec{r}' \rangle}\sigma_{z}^{[\vec{r}]}\sigma_{z}^{[\vec{r}']} - h\sum_{\vec{r}}\sigma_{x}^{[\vec{r}]},
\eea
where $\sigma_{i}^{[\vec{r}]}$ is the $i = (x,z)$ Pauli matrix for site $\vec{r}$, $h$ is the transverse magnetic field, and $\langle \vec{r},\vec{r}' \rangle$ represent nearest-neighbor sites. 
The second example is the spin-1/2 antiferromagnetic Heisenberg model, 
\bea
H = \sum_{\langle \vec{r},\vec{r}' \rangle}\vec{S}^{[\vec{r}]}\vec{S}^{[\vec{r}']},
\eea
where $\vec{S}^{[\vec{r}]} = (\sigma_{x}^{[\vec{r}]},\sigma_{y}^{[\vec{r}]},\sigma_{z}^{[\vec{r}]})/2$. 

In our simulations, we have represented the ground state of the system by a two-site translationally invariant iPEPS made up of two tensors, $A$ and $B$. In order to approximate the ground state of the system,  we have applied imaginary-time evolution together with a second-order Suzuki-Trotter decomposition using $\delta$ down to $10^{-3} -10^{-4}$. To update the tensors, we used the alternating-least-square (ALS) sweep for the subtensors, as explained in Sec.~\ref{sec:reducedupdate}, combined with the FFU. At every update, we have also fixed the gauge of the tensors according to the gauge-fixing described above. Let us stress here that the leading order of the computational cost is the same for both FU and FFU + gauge-fixing schemes, but the prefactor and the subleading correction are different. These turn out to produce a big difference in practical running times, as we shall see. 

In order to assess the advantage of the local gauge fixing for the norm matrix $\tilde{\mathcal{N}}_{LR}$ used in the ALS sweep, we first compare the mean condition number of this matrix between using and not using the gauge. 
As a rule of thumb, the larger the condition number of $\tilde{\mathcal{N}}_{LR}$, the less accuracy we get in solving the system of linear equations at every step of the ALS. The result is shown in Table.~\ref{tab1} for different models and bond dimensions. Overall we see that the condition number of the norm matrix in the case of gauge fixing is improved by several orders of magnitude when compared to the case without gauge fixing. ÊFor completeness, we also shown in the table the condition numbers of matrices $L$ and $R$.  The gauge fixing, thus, improves the stability of the iPEPS algorithm. This result is very similar to what has been obtained for finite PEPS in a similar context.~\cite{Lubasch2}

Besides, we observe that the gauge fixing in the norm matrix also accelerates the convergence in the ALS sweeping. More concretely, in Fig.~\ref{costfunction} we show the convergence of the relative change of the cost function defined in Eq.~(\ref{costfunction_reducedtensors1}), as well as the local fidelity of the subtensors between two iterations $u$ and $u+1$ defined as
\bea
f_{ab}^{u+1}= \frac{( {a_{R}^{u+1}b_{L}^{u+1}} | {a_{R}^{u}b_{L}^{u}})}{\sqrt{({a_{R}^{u+1}b_{L}^{u+1}} | {a_{R}^{u+1}b_{L}^{u+1}})({a_{R}^{u}b_{L}^{u}}|{a_{R}^{u}b_{L}^{u})}}}, 
\eea
where, e.g., $|a_{R}^{u}b_{L}^{u})$, is to be understood as tensors $a_{R}^{u}$ and $b_{L}^{u}$ with their indices reshaped as a vector. We observe that, when the gauge-fixing is applied, both cost function and local fidelity tend to converge faster than the case without gauge fixing, see Fig.~\ref{costfunction}.  
We also noted that speed-up in convergences becomes more significant as the bond dimension $D$ of the iPEPS increases (not shown). 

\begin{figure}
	\includegraphics[width = \columnwidth]{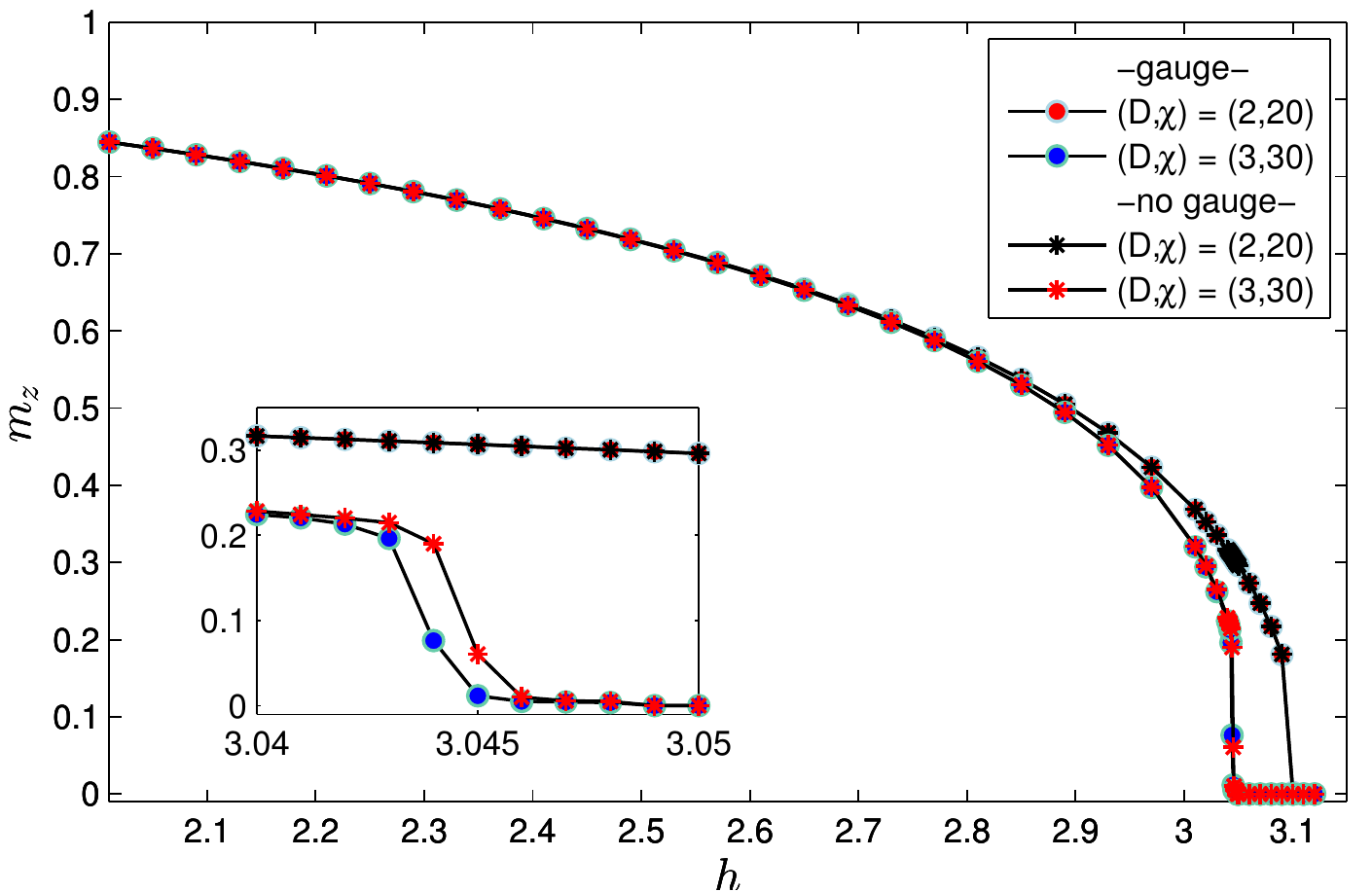}
	\caption{\label{Isingphase} (Color online) Local magnetization $m_z$ as a function of the transverse field $h$ in the quantum Ising model, with gauge (filled circles) and without gauge (star symbols). The inset is a zoom around criticality. }
\end{figure}

\begin{figure}
	\includegraphics[width = \columnwidth]{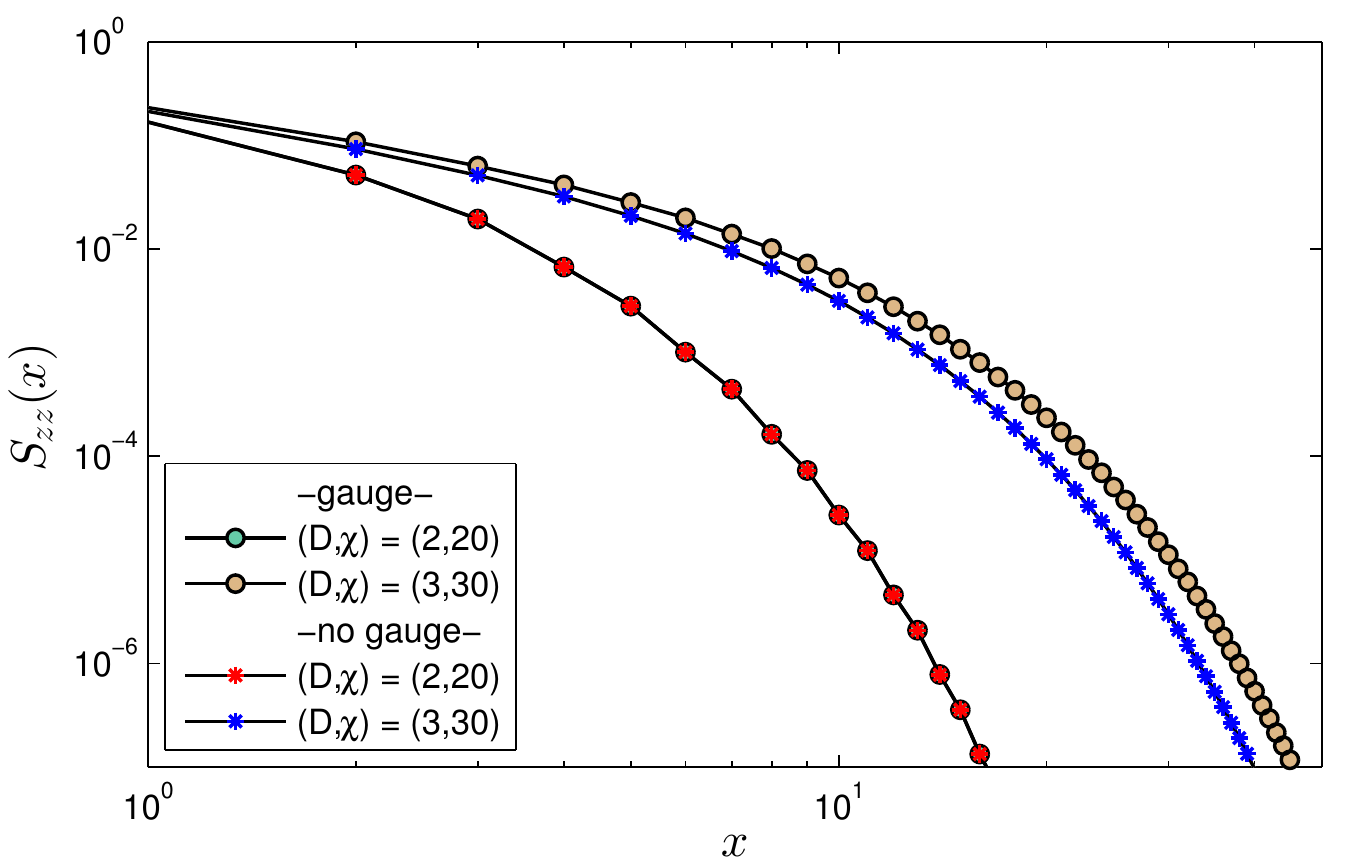}
	\caption{\label{Isingcorr} (Color online) Two-point correlator $S_{zz}(x)$ of the quantum Ising model close to the critical point $h=3.044$,  computed with gauge (filled circles) and without gauge (star symbols).}
\end{figure}

We also compute the order parameter $m_z \equiv \langle \sigma_z \rangle$ for the quantum Ising model, and compare it for the cases with and without gauge fixing, see Fig.~\ref{Isingphase}. The results for both cases are quantitatively very similar deep in the gapped phases. However, when close to the quantum critical point, the results obtained with the iPEPS + gauge-fixing are  better, and again we see that this effect becomes more relevant for larger bond dimensions (e.g., we almost see no difference for $D=2$, whereas for $D=3$ we already see a clear difference). Besides, in Fig.~\ref{Isingcorr} we have plotted the correlation function  $S_{zz}(x) = \langle{\sigma_{z}^l\sigma_{z}^{l+x}}\rangle - \langle{\sigma_{z}^l\rangle\langle\sigma_{z}^{l+x}}\rangle$ at the critical point  $h=3.044$. We see that the simulation with gauge-fixing captures the correlation of the system better  for the case of bond dimensions $(D,\chi) = (3,30)$.

\begin{figure}
	\includegraphics[width = \columnwidth]{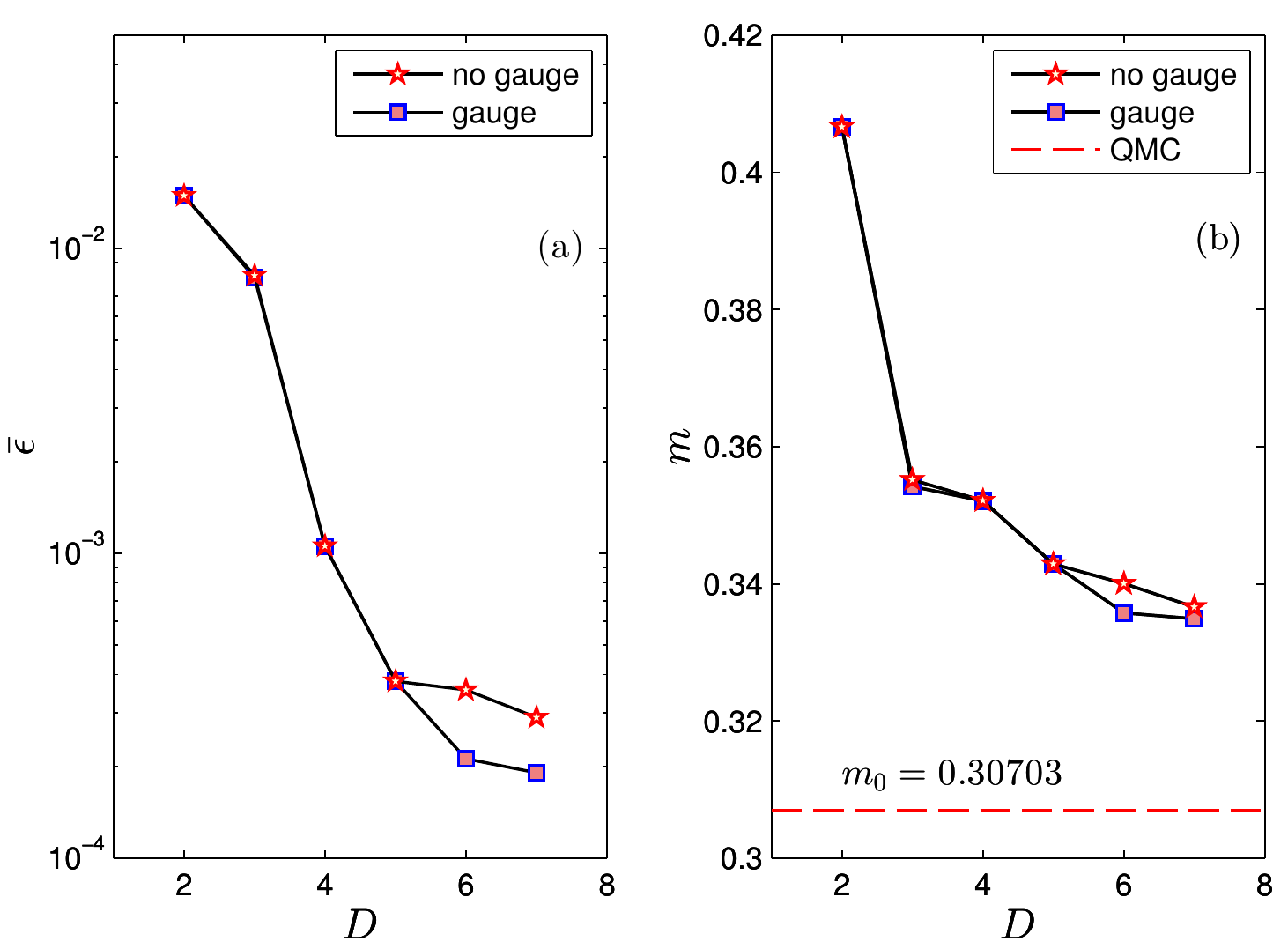}
	\caption{\label{Heis_energy} (Color online) (a) Relative error of the energy per link $\bar{\epsilon} = (\epsilon_0 - \epsilon)/\epsilon_0$ for the Heisenberg model, as a function of the iPEPS bond dimension $D$. Here $\epsilon_0 = -0.669437(5)$ is the quantum Monte Carlo result. \cite{Sandvik_E0} (b) Staggered magnetization $m$ of the Heisenberg model as a function of the bond dimension $D$, compared to the Monte Carlo result $m_{0} = 0.30703$.~\cite{Sandvik_E0}}
\end{figure}

We have also applied the "improved" iPEPS algorithm to study the ground state of the Heisenberg model for different bond dimensions $({D},\chi)$. For small bond dimensions $D$, we do not see much difference for the results with and without gauge fixing. But for large bond dimensions we observe that the ground state energy obtained using gauge fixing is better, see Fig.~\ref{Heis_energy}(a). To quantify the overall error, we compare to the best result obtained from quantum Monte Carlo\cite{Sandvik_E0}, $\epsilon_0 = -0.669437(5)$. In our case, with gauge-fixing we obtain $\epsilon_g = -0.669309(2)$ and without gauge $\epsilon_{ng} = -0.669243(1)$, for $({D},\chi) = (7,70)$. Besides, we also compute the staggered magnetization $m$ as a function of the bond dimension, shown in Fig.~\ref{Heis_energy}(b). Again, for large bond dimensions the calculations with the gauge-fixing become better than without the gauge. Our best values were obtained for $({D},\chi) = (7,70)$, and are $m_{g} = 0.33490$ with gauge $m_{ng} = 0.33662$ without gauge. This is to be compared with the Monte Carlo result, $m_0 = 0.30703$.~\cite{Sandvik_E0} 

\begin{figure}
	\includegraphics[width = \columnwidth]{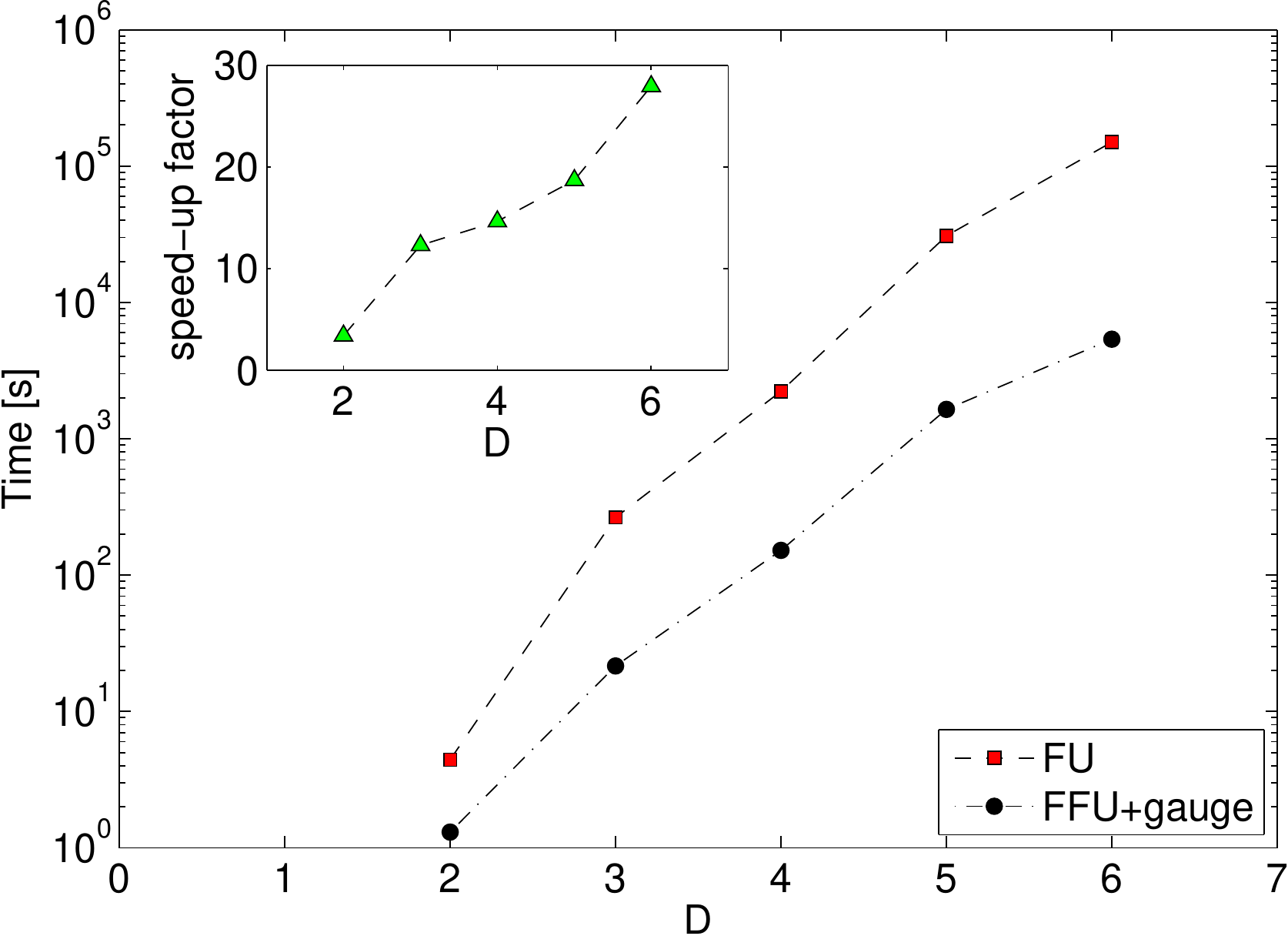}
	\caption{\label{res_fastfull} (Color online) Total running time (in seconds) for five time steps in imaginary-time evolution with the Heisenberg Hamiltonian, with the FU and FFU (+ gauge-fixing) updates. The speed-up factor of the new update scheme with respect to the old one is shown in the inset.}
\end{figure}

Finally, even if not mentioned explicitly at every calculation, we remind that our results are obtained by using the FFU which is remarkably faster than the costly FU update, because the environment is not recomputed from scratch at every step. In Fig.~\ref{res_fastfull} we compare the actual running times for the FU and the FFU (+ gauge-fixing) schemes. The speed-up factor increases with increasing $D$, since more CTM steps are required to reach convergence of the environment tensors if the state is more entangled. In the present example we obtained a speed-up factor of up to $\sim30$. 
This factor will be even larger in more strongly entangled systems (e.g. in fermionic systems). 
In  combination with  the gauge-fixing the FFU approach is even better: we have seen that \emph{the overall improved algorithm is remarkably stable, as well as substantially faster than the old version}. 


One more comment is in order: one could naively expect that after several  updates with the FFU scheme the environment will have drifted away, so that it may need to be
reinitialized by fully converging it. However, we find that the new update scheme is self
correcting for the considered values of $\delta$ and for the studied models (indeed, if $\delta$ is small enough, the changes in the tensors are also expected to be small).  One may expect, however, that for larger $\delta$ the method is no longer self-correcting. This is a possibility that needs to be taken into account when implementing the algorithm in practice. But in any case, $\delta$ decreases throughout the evolution of the algorithm, so the self-correction happens naturally and in practice we never need to restart. For the models analyzed in our paper we never encounter such situation, but we can imagine that for more complex systems one may need to restart the environment from time to time in order to improve the results. This may be also an important difference between imaginary- and real-time evolutions. For imaginary evolutions, the algorithm is naturally self-correcting, since in the limit $\delta \rightarrow 0$ the actual time steps behave, in practice, as convergence steps for the environment. However, this property may be lost for real-time evolutions if $\delta$ is not small enough, so that several iterations may be needed at every step. But in any case, for real-time evolutions it is also important to recycle environment tensors at every iteration in order to save time, as well as to take care of a correct matching of all bond indices as done explicitly in the FFU. 

\section{\label{secV} Conclusions}
In this paper we have presented how to improve the stability and efficiency of the iPEPS algorithm. We have discussed two improvements, namely the fast full update (FFU) and the gauge-fixing in the ALS sweep. In the FFU scheme the tensors in the effective environment are updated at every step, while keeping them compatible with the updated iPEPS tensors. This implies a speed-up by a large factor (up to $\sim30$ in the present examples, and even larger in more strongly entangled systems) compared to the previous FU approach, where the environment tensors have been recomputed from scratch at each time step. The gauge-fixing improves the conditioning in the ALS sweep at every update step, in the same way as was already shown for finite PEPS calculations,~\cite{Lubasch2} leading to a better stability and faster convergence.  

We have benchmarked the improved iPEPS algorithm with calculations for the quantum Ising and Heisenberg models on an infinite 2d square lattice, where we have seen that Êsimilar or slightly better accuracies  can be obtained, substantially faster, and with more stable evolutions, when compared to previous iPEPS calculations. This is particularly true for large bond dimensions and in highly-entangled systems, such as in the vicinity of quantum critical points. 

Technically, we have demonstrated the improved iPEPS algorithm for systems on an infinite square lattice and a 2-site unit cell, but the extension to other 2d lattices and bigger unit cells is straightforward. The method can also be extended to Hamiltonians with longer-range interactions. We expect that these improvements will be a significant step forward towards powerful tensor network calculations for challenging 2d systems. 
\vspace{0.5cm}
\acknowledgments
This research was supported in part by ARC Discovery Projects funding scheme, Project No. DP130104617. R.O. acknowledges funding from the JGU and the DFG. P.C. acknowledges support by the Delta-ITP consortium (a program of the Netherlands Organisation for Scientific Research (NWO) that is funded by the Dutch Ministry of Education, Culture and Science (OCW)).


\begin{thebibliography}{99}
\bibitem{RevTN} R. Or\'us, Annals of Physics {\bf 349} 117-158 (2014); R. Or\'us, Eur. Whys. J. B {\bf 87}, 280 (2014); J. Eisert, Modeling and Sim. 3, 520 (2013); N. Schuch, QIP, Lecture Notes of the 44th IFF Spring School 2013; J. I Cirac, F. Verstraete, J. Phys. A: Math. There. {\bf 42}, 504004 (2009); F. Verstraete, J. I. Cirac, V. Murg, Adv. Phys. {\bf 57}, 143 (2008).
	\bibitem{arealaw1} G. Vidal, J. I. Latorre, E. Rico, A. Kitaev, Phys. Rev.	 Lett. {\bf 90} 227902 (2003)
	 \bibitem{arealaw2} P. Calabrese, J. Cardy, JSTAT 0406:P06002 (2004).
	 \bibitem{arealaw3} M. Srednicki, Phys. Rev. Lett. {\bf 71} 666 (1993).
	 \bibitem{arealaw4} M. Plenio, J. Eisert, J.  Drei\ss{}ig, M. Cramer, Phys. Rev. Lett. {\bf 94}, 060503 (2005)
	\bibitem{arealaw5}	A. Riera, J. I. Latorre, Phys. Rev. A {\bf 74} 052326 (2006)
	\bibitem{Vidal1}G. Vidal, Phys. Rev. Lett. {\bf 91}, 147902 (2003)
	\bibitem{Vidal2}G. Vidal, Phys. Rev. Lett. {\bf 93}, 040502 (2004)
	\bibitem{Frank} D. Poulin, A. Qarry, R. Somma, F. Verstraete, Phys. Rev. Lett. {\bf 106}, 170501 (2011)
	\bibitem{Fannes1}M. Fannes, B. Nachtergaele, and R. F. Werner,Commun. Math. Phys. {\bf 144}, 443 (1992)
	\bibitem{Ostlund1}S. Ostlund and S. Rommer, Phys. Rev. Lett. {\bf 75}, 3537 (1995)
	\bibitem{White1}S. R. White, Phys. Rev. Lett. {\bf 69}, 2863 (1992)
	\bibitem{White2}S. R. White, Phys. Rev. B {\bf 48}, 10345 (1993)
	\bibitem{Takasaki1}H. Takasaki, T. Hikihara, and T. Nishino, J. Phys. Soc. Jpn. {\bf 68}, 1537 (1999)
	\bibitem{schollwock1}U. Schollw\"{o}ck, Rev. Mod. Phys. {\bf 77}, 259 (2005)
	\bibitem{Perez1}D. P\'{e}rez-Garc\'{i}a, F. Verstraete, M.M. Wolf, J.I. Cirac, Quant. Inf. Comput. {\bf 7}, 401 (2007)
	\bibitem{Ian1} I. P. McCulloch, J. Stat. Mech.: Theory Exp. (2007) P10014
	\bibitem{Ian2} I. P. McCulloch, arXiv:0804.2509v1 [cond-mat.str-e1]
	\bibitem{Greg}G. M. Crosswhite, A. C. Doherty and G. Vidal, Phys. Rev. B
	{\bf 78}, 035116 (2008)
	\bibitem{schollwock2}U. Schollw\"{o}ck, Annals of Physics {\bf 326}, 96 (2011).
	\bibitem{tdDMRGWhite} S. R. White and A. E. Feiguin, Phys. Rev. Lett. {\bf 93}, 076401 (2004)
	\bibitem{tdDMRGSchollwock}A. J. Daley, C. Kollath, U. Schollw\"{o}ck, and G. Vidal, J. Stat. Mech.: Theory Exp.{\bf 2004}, P04005 (2004)
	\bibitem{ttn} L. Tagliacozzo, G. Evenbly, G. Vidal, Phys. Rev. B {\bf 80}, 235127 (2009).
	\bibitem{mera} G. Evenbly, G. Vidal, Phys. Rev. Lett. {\bf 102}, 180406 (2009). 
	\bibitem{TDVP}J. Haegeman, J. I. Cirac, T. J. Osborne, I. Pizorn, H. Verschelde, and F. Verstraete, Phys. Rev. Lett. {\bf 107}, 070601 (2011)
	\bibitem{Verstraete1}F. Verstraete and J. I. Cirac, cond-mat/0407066
	\bibitem{Murg1}V. Murg, F. Verstraete and J. I. Cirac, Phys. Rev. A{\bf 75}, 033605 (2007)
	\bibitem{Verstraete2}F. Verstraete, M.M. Wolf, D. P\'{e}rez-Garc\'{i}a, and J.I. Cirac, Phys. Rev. Lett. {\bf 96}, 220601 (2006)
	\bibitem{Nishino2}T. Nishino, K. Okunishi, Y. Hieida, N. Maeshima, and Y. Akutsu, Nucl. Phys. B {\bf 575}, 504 (2000)
	\bibitem{Nishino3}T. Nishino, K. Okunishi, Y. Hieida, N. Maeshima, Y. Akutsu, and A. Gendiar, Prog. Theor. Phys. {\bf 105}, 409 (2001)
	\bibitem{Gendiar1} A. Gendiar, N. Maeshima, and T. Nishino, Pro. Theor. Phys. {\bf 110}, 691 (2003)	
	\bibitem{Levin1} M. Levin and C. P. Nave, Phys. Rev. Lett. {\bf 99}, 120601 (2007)
	\bibitem{Jordan1} J. Jordan, R. Or\'us, G. Vidal, F. Verstraete and J.I. Cirac, Phys. Rev. Lett. {\bf 101}, 250602 (2008)
	\bibitem{Roman2} R. Or\'{u}s and G. Vidal, Phys. Rev. B {\bf 80}, 094403 (2009)
	\bibitem{Jiang1}H. C. Jiang, Z. Y. Weng, and T. Xiang, Phys. Rev. Lett. {\bf 101}, 090603 (2008)
		\bibitem{Xie1}Z. Y. Xie, H. C. Jiang, Q. N. Chen, Z. Y. Weng, and T. Xiang, Phys. Rev. Lett. {\bf 103}, 160601 (2009)	
	\bibitem{Philippe1}P. Corboz, R. Or\'us, B. Bauer, and G. Vidal, Phys. Rev. B {\bf 81}, 165104 (2010)
	\bibitem{Wang1}	L. Wang, I. Pi\v{z}orn, and F. Verstraete, Phys. Rev. B {\bf 83}, 134421 (2011)
	\bibitem{Pizorn1} I. Pi\v{z}orn, L. Wang, and F. Verstraete, Phys. Rev. A {\bf 83}, 052321 (2011)
	\bibitem{Li1} W. Li, J. von Delft, and T. Xiang, Phys. Rev. B {\bf 86}, 195137  (2012)	
	\bibitem{Xie2}	Z. Y. Xie, J. Chen, M. P. Qin, J. W. Zhu, L. P. Yang, and T. Xiang, Phys. Rev. B {\bf 86}, 045139 (2012)
	\bibitem{Romancano1} H. Kalis, D. Klagges, R. Or\'us, K.P. Schmidt, Phys. Rev. A {\bf 86}, 022317 (2012)
	\bibitem{Romancano2} R. Or\'us, H. Kalis, M. Bornemann, K. P. Schmidt, Phys. Rev. A {\bf 87}, 062312 (2013)
	\bibitem{Lubasch1} M. Lubasch, J. Ignacio Cirac, and M.-C. Ba\~{n}uls, New J. Phys. {\bf 16}, 033014 (2014)
	\bibitem{Wang11} L. Wang and F. Verstraete, arXiv:1110.4362 (2011).
	\bibitem{Lubasch2} M. Lubasch, J. Ignacio Cirac, and M.-C. Ba\~{n}uls, Phys. Rev. B {\bf 90}, 064425 (2014)	
	\bibitem{Phiencano1} Ho N. Phien, Ian P. McCulloch, Guifr\'e Vidal, 	arXiv:1411.0391 [quant-ph]
	\bibitem{Bauer} B. Bauer, P. Corboz, R. Or\'us, and M. Troyer, Phys. Rev. B {\bf 83}, 125106 (2011).
	\bibitem{Vidal3}G. Vidal, Phys. Rev. Lett. {\bf 98}, 070201 (2007) 
	\bibitem{Roman1} R. Or\'{u}s and G. Vidal, Phys. Rev. B {\bf 78}, 155117 (2008)
	\bibitem{Bauer1} B. Bauer, G. Vidal and M. Troyer, J. Stat. Mech. P09006 (2009) 
	\bibitem{Philippe3}P. Corboz, S. R. White, G. Vidal and M. Troyer, Phys. Rev. B {\bf 84}, 041108 (2011)
	\bibitem{Philippe5} P. Corboz, T. M. Rice, and M. Troyer, Phys. Rev. Lett. {\bf 113}, 046402 (2014) 
        \bibitem{Philippe5b} P. Corboz and F. Mila, Phys. Rev. B {\bf 87}, 115144 (2013). Y. H. Matsuda, et al., Phys. Rev. Lett. {\bf 111}, 137204 (2013). P. Corboz and F. Mila, Phys. Rev. Lett. {\bf 112}, 147203 (2014).
	\bibitem{Gu} Z.-C. Gu, H.-C. Jiang, D. N. Sheng, H. Yao, L. Balents, and X.-G. Wen, Phys. Rev. B{\bf 88}, 155112 (2013)
	\bibitem{Wang} L. Wang, Z.-C. Gu, F. Verstraete, X.-G. Wen arXiv preprint arXiv:1112.3331 [cond-mat.str-el]
	\bibitem{KHA} M. Ziegler, R. Or\'us, work in progress. 
	\bibitem{Zhao1}H. H. Zhao, Z. Y. Xie, Q. N. Chen, Z. C. Wei, J. W. Cai, and T. Xiang, Phys. Rev. B {\bf 81}, 174411 (2010) 
         \bibitem{RomanNewCTM1} R. Or\'{u}s Phys. Rev. B {\bf 85}, 205117 (2012)
	\bibitem{Baxter11} R. J. Baxter, J. Math. Phys. {\bf 9}, 650 (1968)
	\bibitem{Nishino1}T. Nishino, K. Okunishi, J. Phys. Soc. Jpn. {\bf 65} pp. 891-894 (1996)
	\bibitem{Nishino1_1}T. Nishino, K. Okunishi, J. Phys. Soc. Jpn. {\bf 67} 3066 (1998)
	\bibitem{Philippe4} P. Corboz, J. Jordan, and G. Vidal, Phys. Rev. B {\bf 82}, 245119 (2010)
	\bibitem{GlenRenormT} G. Evenbly, G. Vidal, arXiv:1412.0732 [cond-mat.str-el]
	\bibitem{Frank_corr} F. Verstraete, M. M. Wolf, D. P\'erez-Garc\'ia, J. I. Cirac, Phys. Rev. Lett. {\bf 96}, 220601 (2006)
\bibitem{Suzuki1} M. Suzuki, Phys. Lett. A {\bf 146}, 319 (1990); J. Math. Phys. (N.Y.) {\bf 32}, 400 (1991). 
For higher order expansions, see also A., T. Sornborger and E.â D. Stewart, quant-ph/9809009.
\bibitem{MPOMPDO} G. de las Cuevas, N. Schuch, D. P\'erez-Garc\'ia, J. I. Cirac, New J. Phys. {\bf 15}, 123021 (2013). 
\bibitem{Sandvik_E0} A.â W. Sandvik, Phys. Rev. B {\bf 56}, 11678 (1997)
\end{thebibliography}
\end{document}